\documentclass[11pt,aps,prd,longbibliography]{revtex4-2}
\usepackage{subcaption}
\usepackage{slashed}
\usepackage{graphics}
\usepackage{epsfig}
\usepackage{array}
\usepackage{float}
\usepackage{xcolor}
\usepackage{tikz-feynman}
\usepackage{booktabs}
\usepackage{graphicx}
\graphicspath{{./figures/}}
\usepackage{psfrag}
\usepackage{slashbox}
\usepackage{hyperref}
\usepackage{booktabs}
\usepackage{amsmath}
\usepackage{amssymb}
\usepackage[latin1]{inputenc}
\usepackage{mathrsfs}
\usepackage{xspace}
\usepackage{verbatim}
\newcommand{\UBL}{{\ensuremath{U(1)_{B-L}}}\xspace}

\newcommand{\be}{\begin{equation}\displaystyle}
\newcommand{\ee}{\end{equation}}

\newcommand{\n}{\nonumber\\}

\newcommand{\bes}{\begin{equation*}\displaystyle}
\newcommand{\ees}{\end{equation*}}

\newcommand{\ufo}{{\sc UFO}}

\newcommand{\madgraph}{{\sc MadGraph}}

\newcommand{\madanalysis}{{\sc MadAnalysis}}

\newcommand{\sarah}{{\sc Sarah}}
\newcommand{\spheno}{{\sc SPheno}}
\newcommand{\Higgsbounds}{{\sc HiggsBounds}}
\newcommand{\Higgssignals}{{\sc HiggsSignals}}

\newcommand{\pythia}{{\sc Pythia}}
\newcommand{\delphes}{{\sc Delphes}}

\def\lsim{\raise0.3ex\hbox{$\;<$\kern-0.75em\raise-1.1ex\hbox{$\sim\;$}}}
\def\gsim{\raise0.3ex\hbox{$\;>$\kern-0.75em\raise-1.1ex\hbox{$\sim\;$}}}

\begin{document}
\title{Searching for a Heavy Neutral CP-Even Higgs Boson\\[0.25cm]
 in the BLSSM at the LHC Run 3 and HL-LHC}
\author{M. Ashry$^{1,}$\footnote{mustafa@sci.cu.edu.eg}, S. Khalil$^{2,}$\footnote{skhalil@zewailcity.edu.eg} and S. Moretti$^{3,4,}$\footnote{s.moretti@soton.ac.uk;~stefano.moretti@physics.uu.se}}
\affiliation{
$^1$Department of Mathematics, Faculty of Science, Cairo University, Giza 12613, Egypt\\
$^2$Center for Fundamental Physics, Zewail City of Science and Technology, 6th of October City, Giza 12578, Egypt\\
$^3$School of Physics and Astronomy, University of Southampton, Highfield, Southampton SO17 1BJ, UK\\
$^4$Department of Physics \& Astronomy, Uppsala University, Box 516, SE-751 20 Uppsala, Sweden}

\begin{abstract}
The detection of a heavy neutral CP-even Higgs boson of the $B-L$ Supersymmetric Standard Model (BLSSM), $h'$, with $m_{h'}\simeq 400~\text{GeV}$, at the Large Hadron Collider (LHC) for a center-of-mass energy of $\sqrt{s}=14~\text{TeV}$, is investigated. The following production and decay channels are considered: $gg\to h'\to{ZZ}\to4\ell$ and $gg\to h'\to{W^+W^-}\to2\ell+\slashed{E}_T$ (with $\slashed{E}_T$ being the 
Missing~Transverse~Energy~(MET)), where $\ell=e,\mu$, with integrated luminosity $L_{\text{int}}=300~{\text{fb}}^{-1}$ (Run 3). Furthermore, we also look into the di-Higgs channel $gg\to h'\to{hh}\to{b\bar{b}\gamma\gamma}$ at the High-Luminosity LHC (HL-LHC) with an integrated luminosity of $L_{\text{int}}=3000~{\text{fb}}^{-1}$.
{We demonstrate that promising signals with high signal-to-background statistical significance ($S/\sqrt{B}$) can be obtained through the three aforementioned channels.}
\end{abstract}
\maketitle

\section{\label{Sec:ntrdctn}INTRODUCTION}

The search for a heavy neutral CP-even Higgs boson at the current Run 3 of the LHC and a future HL-LHC is an active area of research~\cite{ATLAS:2022eap,ATLAS:2022rws,CMS:2022pjv,Adhikary:2020ujn,Chen:2019hey,Bahl:2020kwe,Gu:2017ckc,Banerjee:2015hoa,Hammad:2016trm}. 
This is so because virtually any extension of the Higgs sector beyond the single doublet structure of the Standard Model (SM), in which the only neutral CP-even state of it is identified with the particle that was discovered in 2012 at the LHC by the ATLAS and CMS experiments~\cite{ATLAS:2012yve,CMS:2012qbp}, contains it. As a result, currently, probing such a heavy Higgs boson is one of the main goals of the LHC experiments, as it could well provide the first hint for physics Beyond the SM (BSM). 
Both ATLAS and CMS have searched for a heavy Higgs boson and the corresponding analyses typically involve looking for events in which the heavy Higgs boson is produced and then decays into SM particles, such as $W^\pm$ or $Z$ bosons, in turn decaying into leptons or jets~\cite{ATLAS:2022eap}, or into the SM Higgs boson itself~\cite{Chernyavskaya:2023uqg}, which then decays into, e.g., photons, $b$-quarks or $ \tau$ leptons.

Supersymmetric extensions of the SM are one of the BSM frameworks that consistently predict the existence of several Higgs bosons, including a heavy neutral CP-even one. Such a Higgs boson mass can be significantly larger than the one of the SM Higgs state, potentially reaching several hundred GeV. For example, the Minimal Supersymmetric Standard Model (MSSM) contains five Higgs bosons: two CP-even ($h$ and $H$, with $m_h<m_H$), one CP-odd ($A$) and two charged states ($H^+$ and $H^-$): for reviews, see, e.g.,~\cite{Djouadi:2005gj}. This is the simplest construct implementing supersymmetry, where the 
 lightest CP-even Higgs boson, $h$, is designated as the SM Higgs boson, with a mass of 125 GeV, which, however, imposes a strenuous configuration on the MSSM parameter space, forcing the other CP-even Higgs boson, $H$, to be rather heavy and significantly decoupled. 
However, if supersymmetry is non-minimal, in either its gauge or Higgs sector or both,
then the mass of additional CP-even Higgs states can become rather less constrained~\cite{Moretti:2019ulc}. An example of
this is the so-called BLSSM, which indeed offers the possibility of LHC signals for a CP-even Higgs state not only above the SM Higgs mass, e.g., in the range up to 500 GeV~\cite{Hammad:2016trm}, but also afford one with a lighter mass spectrum, in turn able to explain past~\cite{Hammad:2015eca,Abdallah:2014fra} and present data anomalies~\cite{Abdelalim:2020xfk}.

The BLSSM is a theoretical extension of the MSSM that includes an additional $U(1)$ gauge symmetry known as $B-L$ (baryon number minus lepton number)~\cite{Khalil:2007dr,OLeary:2011vlq,Basso:2011hn,Basso:2015xna} as well as an extended Higgs sector. The $B-L$ symmetry is motivated by the observation that the difference between baryon and lepton number is conserved in many particle physics processes. In the BLSSM, the $B-L$ symmetry may be broken at the few TeV scale, giving rise to new particles such as two new extra neutral CP-even Higgs bosons. One of them, labeled $h'$, can have energies in the hundreds of GeV range. It is indeed the presence of such a $h'$ state that causes the aforementioned new phenomenology to emerge in collider experiments, which can then be used to test the BLSSM hypothesis.

We emphasize that the SM-like Higgs state, henceforth labeled by $h$, is derived from the real parts of the neutral components of the Electro-Weak (EW) scalar doublets $H_u$ and $H_d$ whereas the (typically) next-to-lightest Higgs boson, $h'$, stems from the real parts of the neutral components of the $B-L$ scalar singlets $\chi_1$ and $\chi_2$. Despite the fact that the mass mixing between these two types of Higgs bosons is negligible, a non-vanishing kinetic mixing allows for relevant couplings between $h'$ and the SM particles, resulting in a total cross section of $h'$ production and decay into $W^+W^-$, $ZZ$ or $hh$ of $\mathcal{O}(1)~{\text{fb}}$. These signals are typically smaller than the associated backgrounds but, by using appropriate selection strategies, they can be probed with a reasonably high sensitivity. However, given that current
experimental limits have significantly constrained also the BLSSM parameter space above and beyond what allowed for in 
Ref.~\cite{Hammad:2016trm}, which targeted Run 2 sensitivities, we revisit here the scope of Run 3 and the HL-LHC in accessing the heavy neutral CP-even Higgs boson of the BLSSM, $h'$, in the mass region of 400 GeV or so. 
It is also worth mentioning that heavy Higgs boson searches have been conducted in many supersymmetric (and non-supersymmetric) extensions of the SM. Indeed, the BLSSM itself has been phenomenologyically investigated rather widely in relation to Higgs, dark matter and heavy gauge boson physics due to its many degrees of freedom and its wide parameter space~\cite{Antusch:2021oit,Ezzat:2021bzs,Ashry:2021qrt,Bhattacharyya:2023ehx,Abdallah:2018kix,Abdallah:2014fra,Abdallah:2016vcn,Hammad:2016trm,Hammad:2015eca,Dev:2023kzu,Un:2016hji,Abdelalim:2020xfk,Boubaa:2022xsk,Elsayed:2011de,Basso:2010jm}. Specifically,
for heavy Higgs bosons, the situation in the BLSSM is significantly different from that of the MSSM, where the SM-like Higgs boson mass and couplings constrain the heavy Higgs boson phenomenology greatly. In contrast, in the BLSSM, while the Higgs bosons of MSSM origin are just as restricted as in the actual minimal model, the constraints on the other Higgs bosons from the $B-L$ sector are much relaxed in comparison.

The paper is organized as follows. We briefly review the BLSSM particle content, superpotential and gauge structure in Sec.~\ref{Sec:hgsblssm}, where we also discuss at some length its Higgs sector. Studies of $h'$ signals at the LHC are then carried out in Sec.~\ref{Sec:srchH}, wherein a detailed Monte Carlo (MC) analysis for $h'$ production via (mostly) gluon-gluon fusion (ggF) and decay via $W^+W^-\to2\ell+\slashed{E}_T$, $ZZ\to4\ell$ and $hh\to $
$b\bar{b}\gamma\gamma$ is performed. Our conclusions and final remarks are given in Sec.~\ref{Sec:cnclsn}.

\section{\label{Sec:hgsblssm}The BLSSM}
The BLSSM is based on the gauge symmetry group $SU(3)_C\otimes SU(2)_L\otimes U(1)_Y\otimes U(1)_{B-L}$. This model is a natural extension of the MSSM, with: i) three chiral singlet superfields $\hat{N}_i$ introduced to cancel the $U(1)_{B-L}$ triangle anomaly and acting as right-handed neutrinos, thereby accounting for the measurements of light neutrino masses; ii) two chiral SM-singlet Higgs superfields ($\hat{\chi}_1, \hat{\chi}_2$) with $B-L$ charge $=\pm 2$ to spontaneously break the $U(1)_{B-L}$ gauge group; iii) a vector superfield, $Z'$, necessary to gauge $U(1)_{B-L}$. 
The quantum numbers of the chiral superfields with respect to the SM gauge group ($\mathbb{G}_{\text{SM}}=SU(3)_C\otimes SU(2)_L\otimes U(1)_Y$) and the $\UBL$ one are summarized in Table~\ref{Tab:blssmsf}, {where the $U(1)_{Y,B-L}$ charges generators are given by $Q_{Y}=Y/2,~Q_{BL}=B-L$ and the covariant derivative is $D_\mu\supset-i\big[g_1Q_YV_\mu+(\tilde{g}Q_Y+g_{BL}Q_{BL})V'_\mu\big]$, where $\tilde{g}$ is the gauge kinetic mixing, as discussed below, with $V_\mu$ and $V'_\mu$ are the $U(1)_Y$ being the $U(1)_{B-L}$ gauge fields, respectively.}
\begin{table}[ht]
\captionsetup{justification=centering}\centering
\begin{center}
	\begin{tabular}{cccccc} 
		\hline\hline 
		Superfield & Spin-0 & Spin-$\frac{1}{2}$ & Generations & $\mathbb{G}_{\text{SM}}\otimes \UBL$ \\ 
		\hline 
		$\hat{Q}$ & $\tilde{Q}$ & $Q$ & 3 & $\big({\bf 3},{\bf 2},~~\frac{1}{6},~~\frac{1}{3}\big)$ \\ 
		$\hat{d}^c$ & $\tilde{d}^c$ & $d^c$ & 3 & $\big({\bf \overline{3}},{\bf 1},~~\frac{1}{3},-\frac{1}{3}\big)$ \\ 
		$\hat{u}^c$ & $\tilde{u}^c$ & $u^c$ & 3 & $\big({\bf \overline{3}},{\bf 1},-\frac{2}{3},-\frac{1}{3}\big)$ \\ 
		$\hat{L}$ & $\tilde{L}$ & $L$ & 3 & $\big({\bf 1},{\bf 2},-\frac{1}{2},-1\big)$ \\ 
		$\hat{E}^c$ & $\tilde{e}^c$ & $e^c$ & 3 & $\big({\bf 1},{\bf 1},~~1,~~1\big)$ \\ 
		$\hat{N}^c$ & $\tilde{N}^c$ & $N^c$ & 3 & $\big({\bf 1},{\bf 1},~~0,~~1\big)$ \\ 
		$\hat{H}_d$ & $H_d$ & $\tilde{H}_d$ & 1 & $\big({\bf 1},{\bf 2},-\frac{1}{2},~~0\big)$ \\ 
		$\hat{H}_u$ & $H_u$ & $\tilde{H}_u$ & 1 & $\big({\bf 1},{\bf 2},~~\frac{1}{2},~~0\big)$ \\ 
		$\hat{\chi}_1$ & $\chi_1$ & $\tilde{\chi}_1$ & 1 & $\big({\bf 1},{\bf 1},~~0,-2\big)$ \\ 
		$\hat{\chi}_2$ & $\chi_2$ & $\tilde{\chi}_2$ & 1 & $\big({\bf 1},{\bf 1},~~0,~~2\big)$ \\ 
		\hline \hline
	\end{tabular} 
\caption{\label{Tab:blssmsf}Chiral superfields and their quantum numbers in the BLSSM.}
\end{center}
\end{table}

The BLSSM superpotential is given by 
\begin{equation}
W = Y^{ij}_u \hat{u}^c_i \hat{Q}_j\cdot\hat{H}_u 
- Y^{ij}_d \hat{d}^c_i \hat{Q}_j\cdot\hat{H}_d 
- Y^{ij}_e \hat{E}^c_i \hat{L}_j\cdot\hat{H}_d 
+Y^{ij}_{\nu} \hat{N}^c_i \hat{L}_j\cdot\hat{H}_u 
+\frac{1}{2}Y^{ij}_{N} \hat{N}^c_i \hat{\chi}_1 \hat{N}^c_j 
+ \mu \hat{H}_u\cdot\hat{H}_d 
- \mu' \hat{\chi}_1 \hat{\chi}_2. 
\end{equation}
{
The relevant soft supersymmetry-breaking terms, adopting the usual
universality assumptions at the Grand Unification Theory (GUT) scale, are given by
\begin{align}
-\mathcal{L}_{\text{soft}}
&=m_0^{2}\sum_{\phi} \vert \phi \vert^2 +Y^A_u\tilde{Q}H_u\tilde{U}^c + Y^A_d \tilde{Q}H_d\tilde{D}^c+ Y^A_e\tilde{L}H_d\tilde{E}^c+Y_{\nu}^{A}\tilde{L}H_u\tilde{\nu}^{c}+Y_{N}^{A}\tilde{N}^{c}\chi_1\tilde{N}^{c}\n
&+\Big[B \left(\mu H_u H_d + \mu' \chi_1 \chi_2\right)+\frac{1}{2}m_{1/2}\left(\tilde{g}^a \tilde{g}^a+\tilde{W}^a \tilde{W}^a+ \tilde{B}\tilde{B} + \tilde{B'}\tilde{B'}\right)+ h.c.\Big],
\label{soft}
\end{align}
where the sum in the first term runs over the scalar fields $\phi=\tilde{Q},\tilde{U},\tilde{D},\tilde{L},\tilde{E},\tilde{N},H_{u,d},\chi_{1,2}$
and $(Y_f^A)_{ij}\equiv A_0 (Y_f)_{ij}$ ($f=u,d,e,\nu,N$) are the trilinear scalar interaction couplings associated with the fermion Yukawa couplings. The $B-L$ symmetry can be radiatively broken by
the following non-vanishing Vacuum Expectation Values (VEVs): $\langle
\chi_1 \rangle = v_1$ and $\langle \chi_2 \rangle = v_2$. We define $\tan{\beta'}$ as the ratio of these VEVs ($\tan{\beta'}=v_1/v_2$) in analogy to the MSSM case ($\tan{\beta}=v_u/v_d$)~\cite{Khalil:2016lgy,Khalil:2007dr}.
}

After $B-L$ Spontaneous Symmetry Breaking (SSB), the new gauge boson, $Z'$, acquires its mass from the kinetic term of the $B-L$ Higgs fields, $\chi_{1,2}$. Namely, we have
\be
M_{Z'}^2 = g_{BL}^2 v'^2 + \frac{1}{4} \tilde{g}^2 v^2 ,
\ee
where $\tilde{g}$ is the gauge coupling mixing between $U(1)_Y$ and $U(1)_{B-L}$ and $v'=\sqrt{v_1^2+v_2^2}$. Furthermore, the mixing angle between the (SM) $Z$ and (BLSSM) $Z'$ states is given by 
\be 
\tan 2 \theta' = \frac{2 \tilde{g}\sqrt{g_1^2+g_2^2}}{\tilde{g}^2 + 16 (\frac{v'}{v})^2 g_{BL}^2 -g_2^2 -g_1^2},
\ee
which should be $\lsim 10^{-3}$. 

We now turn to the neutral CP-even Higgs bosons in the BLSSM. The Higgs potential is
\begin{align} 
V(H,\chi)&=|\mu|^2 (|H_u^0|^2+|H_d^0|^2)+|\mu'|^2 (|\chi_1|^2+|\chi_2|^2)+\frac{g^2}{8}(|H_u^0|^2-|H_d^0|^2)^2
+\frac{g_{BL}^2}{2} (|\chi_1|^2-|\chi_2|^2)^2\n
&-\frac{\tilde{g}g_{BL}}{4} (|H_u^0|^2-|H_d^0|^2)(|\chi_1|^2-|\chi_2|^2)
- m_1^2 |\chi_1|^2 - m_2^2 |\chi_2|^2 - B'_{\mu} \chi_1 \chi_2.
\label{Eq:Higgspot}
\end{align} 
where $g^2=g_1^2+g_2^2+\tilde{g}^2$. We expand the neutral components around their VEVs:
\begin{equation}
H_{u,d}^0=\frac{1}{\sqrt{2}}(v_{u,d}+\sigma_{u,d}+i\phi_{u,d}),\quad 
\chi_{1,2}=\frac{1}{\sqrt{2}}(v_{1,2}+\sigma_{1,2}+i\phi_{1,2}).
\end{equation}
The Higgs bosons (symmetric) mass matrix in the basis $(\sigma_u,\sigma_d,\sigma_1,\sigma_2)$ is given in block form by 
\begin{equation}
M_H^2=\begin{pmatrix} M_{HH}^2 & M_{H\chi}^2\\ . & M_{\chi\chi}^2 \end{pmatrix}, 
\end{equation}
where the off-diagonal block mixing of both the MSSM and $B-L$ sectors is
\begin{equation}
M_{H\chi}^2
=\frac{vv'}{2}\tilde{g}g_{BL}\begin{pmatrix} -s_\beta s_{\beta'} & ~~s_\beta c_{\beta'} \\ ~~c_\beta s_{\beta'} & -c_\beta c_{\beta'} \end{pmatrix} 
\end{equation}
where we have used the shorthand notations $s_X\equiv\sin X$ and $c_X\equiv\cos X$.
The MSSM Higgs mass matrix $M_{HH}^2$ in the basis $(\sigma_u,\sigma_d)$ is given by 
\begin{align}
M_{HH}^2&=\begin{pmatrix}
 \frac{\tilde{g}g_{BL}}{4}v'^2 c_{2\beta'}+\frac{1}{8} g^2v^2(23s_\beta^2-7c_\beta^2)+\frac{B_\mu}{t_\beta} & -B_{\mu}-\frac{g^2}{4}v^2s_{2\beta} \\
 . & -\frac{\tilde{g}g_{BL}}{4}v'^2c_{2\beta'}-\frac{1}{8} g^2v^2(7s_\beta^2-23c_\beta^2)+B_\mu t_\beta
\end{pmatrix},
\end{align}
where we have used the shorthand notation $t_X\equiv\tan X$ 
and the $B-L$ Higgs mass matrix $M_{\chi\chi}^2$ in the basis $(\sigma_1,\sigma_2)$ is given by
\begin{align}
M_{\chi\chi}^2&=\begin{pmatrix}
 \frac{\tilde{g}g_{BL}}{4}v^2c_{2\beta}+\frac{1}{2} g_{BL}^2v'^2(23s_{\beta'}^2-7c_{\beta'}^2)+\frac{B'_\mu}{t_{\beta'}} & -B'_{\mu}- g_{BL}^2 v'^2s_{2\beta'} \\
 . & -\frac{\tilde{g}g_{BL}}{4}v^2c_{2\beta}-\frac{1}{2} g_{BL}^2v'^2(7s_{\beta'}^2-23c_{\beta'}^2)+B'_\mu t_{\beta'}
\end{pmatrix},
\end{align}
where the tree-level tadpole equations solutions give
\begin{align}
B_\mu&=-\frac{1}{8}t_{2\beta}\big[g^2v^2c_{2\beta}-2g_{BL} \tilde{g} v'^2 c_{2\beta'}+4(m_d^2-m_u^2)\big],\n
B'_\mu&=-\frac{1}{4}t_{2\beta'}\big[2g_{BL}^2v'^2c_{2\beta'}-g_{BL} \tilde{g} v^2 c_{2\beta}+2(m_2^2-m_1^2)\big],
\end{align}
{where $m_{u,d}^2,~m_{1,2}^2$ are the soft supersymmetry breaking Higgs $H_{u,d},\chi_{1,2}$ mass parameters at the SSB scale(s).}

\begin{figure}[!t]
\centering
\includegraphics[scale=0.8]{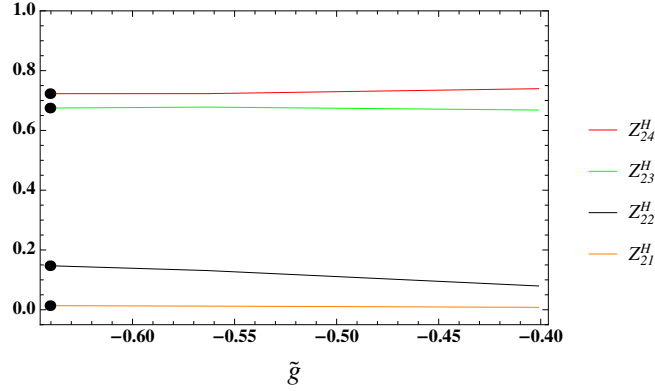}
\caption{\label{Fig:mhpgtgbl}The Higgs mixing $Z^H_{2i}~(i=1,\ldots,4$) versus the gauge kinetic mixing coupling $\tilde{g}$. The values corresponding to the Benchmark Point (BP) of (forthcoming) Table~\ref{Tab:BPPar} are labeled by $\bullet$.}
\end{figure}

The heavy Higgs boson tree-level mass eigenvalues are given in terms of the lightest SM-like Higgs boson $h\equiv{h_1}$ mass, which is fixed at $m_{h}=125~\text{GeV}$, and the lightest $B-L$ Higgs boson $h'\equiv{h_2}$ mass, which we take to be $m_{h'}=400~\text{GeV}$, as follows
\begin{equation}
m_{H,H'}^2=\frac{1}{2}\big(T_H-m_{h}^2-m_{h'}^2\big)\left[1\pm\sqrt{1-\frac{4D_H}{m_{h}^2 m_{h'}^2(T_H-m_{h}^2-m_{h'}^2)^2}}~\right],
\end{equation}
where $D_H=\text{Det}(M_H^2)$, and the trace $T_H=\text{Tr}(M_H^2)$ is given by
\begin{equation}
T_H=2|\mu|^2+2|\mu'|^2+m_u^2+m_d^2+m_1^2+m_2^2+2g^2v^2+8g_{BL}^2v'^2.
\end{equation}
For $\sigma_i=\sigma_d,\sigma_u,\sigma_1,\sigma_2$, one has $\sigma_i=Z^H_{ji}h_j,~h_j=h,h',H,H'$ and conversely $h_j=Z^H_{ij}\sigma_i$. Further, 
\begin{align}
h'&\approx Z^H_{22}\sigma_d+Z^H_{23}\sigma_1+Z^H_{24}\sigma_2.
\end{align}

In Fig.~\ref{Fig:mhpgtgbl}, we display the mixing $Z^H_{2i}$ versus the gauge kinetic mixing $\tilde{g}$. As it can be seen from this plot, $h'$ is essentially generated from $\sigma_{1,2}$ with smaller contributions from the real components of $\sigma_d$ which, however, connect it to the SM sector. 
The MSSM gauginos (bino, wino and gluino) soft masses are fixed to $M_{\tilde{B}}\sim7.74\times10^2~{\text{GeV}},~M_{\tilde{W}}\sim8.52\times10^2~\text{GeV}$ and $M_{\tilde{g}}\sim6.38\times10^2~\text{GeV}$ at the SSB scale(s), respectively, while the $B-L$ gaugino (bino$'$) soft mass $M_{\tilde{B}'}$, and the bino-bino$'$ gauginos mixing soft mass $M_{\tilde{B}\tilde{B}'}$ are given in Table~\ref{Tab:BPPar}.
\begin{table}[!t]
\begin{subtable}{1.\textwidth}
\captionsetup{justification=centering}\centering
{
\begin{tabular}{cccccccccccccccc}
\hline\hline
$g_{BL}$	& $\tilde{g}$	& $t_{\beta}$	& $t_{\beta'}$	& $v'$	& $m_u^2$			& $m_d^2$			& $m_1^2$			& $m_2^2$			& $M_{\tilde{B}'}$	& $M_{\tilde{B}\tilde{B}'}$ \\\hline
$0.675$		& $-0.640$ 		& $11.034$		& $1.288$		& $4875$& $-1.30\times10^7$ & $9.30\times10^6$	& $-5.75\times10^5$ & $4.02\times10^6$& $1.49\times10^3$	& $-1.55\times10^3$ \\
\hline\hline
\end{tabular}
\caption{\label{Tab:BPPar}BP inputs (mass parameters are in GeV.)}
}
\end{subtable}
\\\vspace{0.3cm}
\begin{subtable}{1.0\linewidth}
\captionsetup{justification=centering}\centering
{
\begin{tabular}{ccccccccccccccccc} 
\hline\hline
$Z^H_{11}$	& $Z^H_{12}$	& $Z^H_{13}$	& $Z^H_{14}$	& $Z^H_{21}$	& $Z^H_{22}$	& $Z^H_{23}$ 	& $Z^H_{24}$	& $Z^H_{31}$	& $Z^H_{32}$	& $Z^H_{33}$	& $Z^H_{34}$	& $Z^H_{41}$	& $Z^H_{42}$ 	&	$Z^H_{43}$ 	&	$Z^H_{44}$ \\\hline
$0.089$ 	& $0.987$ 		& $-0.100$ 		& $-0.088$ 		& $0.012$ 		& $0.131$ 		& $0.678$		& $0.723$ 		& $0.030$		& $0.009$ 		& $0.728$ 		& $-0.685$ 		& $-0.995$ 		& $0.091$		& $0.021$ 		& $-0.019$\\
\hline\hline
\end{tabular}
\caption{\label{Tab:BPHMx}BP neutral CP-even Higgs mixing.}
}
\end{subtable}
\\\vspace{0.3cm}
\begin{subtable}{1.0\linewidth}
\captionsetup{justification=centering}\centering
{
\begin{tabular}{cccccccc}
\hline\hline
$m_{H^{\pm}}$ 	& $m_{A}$	& $m_{A'}$	& $m_{h_1\equiv h}$	& $m_{h_2\equiv h'}$	& $m_{h_3\equiv H'}$	& $m_{h_4\equiv H}$	& $M_{Z'}$ \\\hline
$4384$ 			& $2587$ 	& $4384$ 	& $125$ 			& $397$ 				& $4241$ 				& $4402$ 			& $3300$ \\
\hline\hline
\end{tabular}
\caption{\label{Tab:BPSpc}BP Higgs mass spectrum and $M_{Z'}$ (GeV).}
}
\end{subtable}
\captionsetup{justification=centering}\centering
{\caption{\label{Tab:BP}BP and relevant outputs.}}
\end{table}

The second lightest Higgs boson $h'$ interaction couplings to quarks are given in terms of quark masses $M_{u,d}$ by
\begin{align}
\label{Eq:ghff}
\Gamma^{h'}_{\bar{u}u}=-\frac{M_u}{v s_\beta} Z^{H}_{22},\quad\quad
\Gamma^{h'}_{\bar{d}d}=-\frac{M_d}{v c_\beta} Z^{H}_{22},
\end{align}
while its couplings to the SM gauge and Higgs bosons are given by
\begin{align}
\label{Eq:ghww}
g_{h'WW}&\approx g_2 M_W s_\beta Z^H_{22},\\
\label{Eq:ghzz}
g_{h'ZZ}&\approx g_{h'WW} \Big(\sec\theta_w - \frac{\tilde{g}}{g_2} s_{\theta_{w'}}\Big)^{2},\\
g_{h'hh}&\approx
\frac{1}{4}Z_{{22}}^{H} \Big[4 \tilde{g} g_{BL} v' Z_{{12}}^{H} \Big(Z_{{13}}^{H} s_{\beta'}- Z_{{14}}^{H} c_{\beta'}\Big)-3g^2 v s_{\beta} (Z_{{12}}^{H})^2\Big]\n
&+\frac{1}{2}Z_{{23}}^{H} \Big[2 g_{BL}^{2} v' \Big(3 s_{\beta'} (Z_{{13}}^{H})^2 -2 c_{\beta'} Z_{{13}}^{H} Z_{{14}}^{H} - s_{\beta'} (Z_{{14}}^{H})^2\Big)
-\tilde{g} g_{BL} \Big(v' s_{\beta'} (Z_{{12}}^{H})^2+2v s_{\beta} Z_{{12}}^{H} Z_{{13}}^{H}\Big)\Big]\n
&+\frac{1}{2}Z_{{24}}^{H} \Big[2 g_{BL}^{2} v' \Big(c_{\beta'} (Z_{{13}}^{H})^2 +2 s_{\beta'} Z_{{13}}^{H} Z_{{14}}^{H}-3 c_{\beta'} (Z_{{14}}^{H})^2\Big)\Big],
\label{Eq:ghhh}
\end{align}
where $\theta_w$ and $\theta_{w'}$ are the weak and $Z-Z'$ mixing angles, respectively.
For $t_{\beta'}\sim1$, $Z_{{12}}^{H}\sim1,~Z_{{13}}^{H},Z_{{14}}^{H}\ll1,~Z_{{23}}^{H},Z_{{24}}^{H}\sim\frac{1}{\sqrt{2}}$, the trilinear Higgs boson coupling $h' hh$ (relevant to our forthcoming analysis) is approximated by 
\begin{align}
g_{h'hh}
&\sim -\frac{1}{2} \Big( \frac{3}{2} g^2 v s_{\beta} Z_{{22}}^{H} +\tilde{g} g_{BL} v' \Big).
\label{Eq:ghhh1}
\end{align}
%

\section{\label{Sec:srchH}SEARCH FOR A HEAVY Neutral CP-EVEN Higgs BOSON AT THE LHC}

Many computational tools are used throughout this work, from building the model analytically to performing the numerical simulations at detector level. The BLSSM was first implemented into the \sarah~package for Mathematica and the output was then passed to \spheno~\cite{Staub:2013tta,Porod:2011nf} for numerical calculations of the particle spectrum. After that, the ensuing \ufo~model was used in \madgraph~\cite{Alwall:2011uj} for MC event generation and Matrix Element (ME) calculations. After that, \pythia~was used to simulate initial and final state radiation (through the Parton Shower (PS) formalism) 
as well as fragmentation/hadronization effects~\cite{Sjostrand:2007gs}. For detector simulation, the \pythia~output was passed to \delphes~\cite{deFavereau:2013fsa}. Finally, for data analysis, we used \madanalysis~\cite{Conte:2012fm}. As for the BP used, we made sure that it was consistent
with \Higgsbounds~and \Higgssignals~\cite{Bechtle:2013wla,Bechtle:2013xfa} limits, as obtained from the latest LHC data. 

\begin{figure}[!t]
\centering
\includegraphics[trim={4.5cm 20.5cm 4cm 4cm},clip,scale=.88]{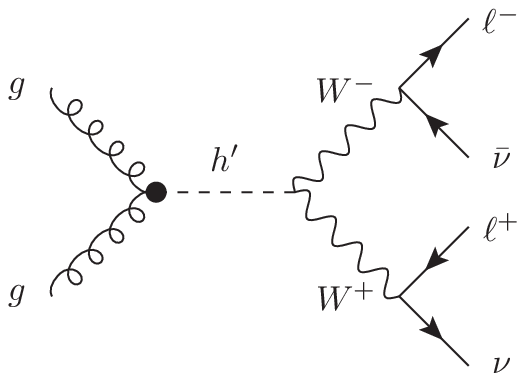}~\hspace{-3.2in}~
\includegraphics[trim={2cm 20.5cm 4cm 4cm},clip,scale=.88]{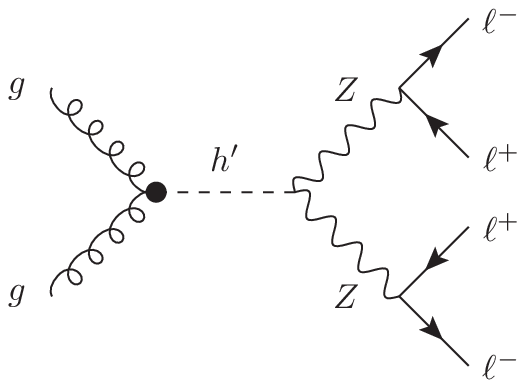}~\hspace{-3.2in}~
\includegraphics[trim={2cm 20.5cm 4cm 4cm},clip,scale=.88]{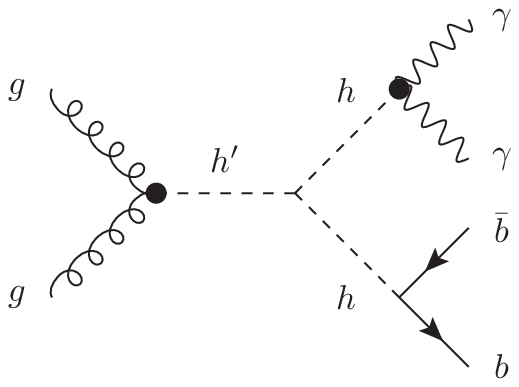}
\caption{\label{Fig:feynmanHhhVV}Feynman diagrams for $h'$ production via ggF and decays via (from left to right) $W^+W^-\to 2\ell+\slashed{E}_T$, $h'\to ZZ\to 4\ell$ and $h'\to{hh}\to{b\bar{b}\gamma\gamma}$.}
\end{figure}
The Feynman diagrams associated to the $h'$ production and decay mechanisms discussed here are found in Fig.~\ref{Fig:feynmanHhhVV}, wherein the $\bullet$ symbol is meant to signify the exact loop function allowing for both $b$ and $t$ quark contributions. The Higgs production and decay rates are computed by factorising the $h'$ propagator, so that the overall event yield can be broken down into the $h'$ production cross section and decay Branching Ratios (BRs).
The MC event generation is done at Leading Order (LO) for both Signal ($S$) and Background ($B$), however, we include Next-to-Next-to-LO (NNLO) inclusive $k$-factors from Quantum Chromo-Dynamics (QCD) in computing our significances, specifically, we use 2.2 for the ggF signal and 1.2 for the Vector Boson Fusion (VBF) one (see below) as well as the (EW) backgrounds~\cite{Harlander:2001eb,PhysRevLett.88.201801,Ciccolini:2007ec,Baglio:2018lrj,Schmidt:2015cea}.
\begin{figure}[!t]
\centering
\includegraphics[scale=.40]{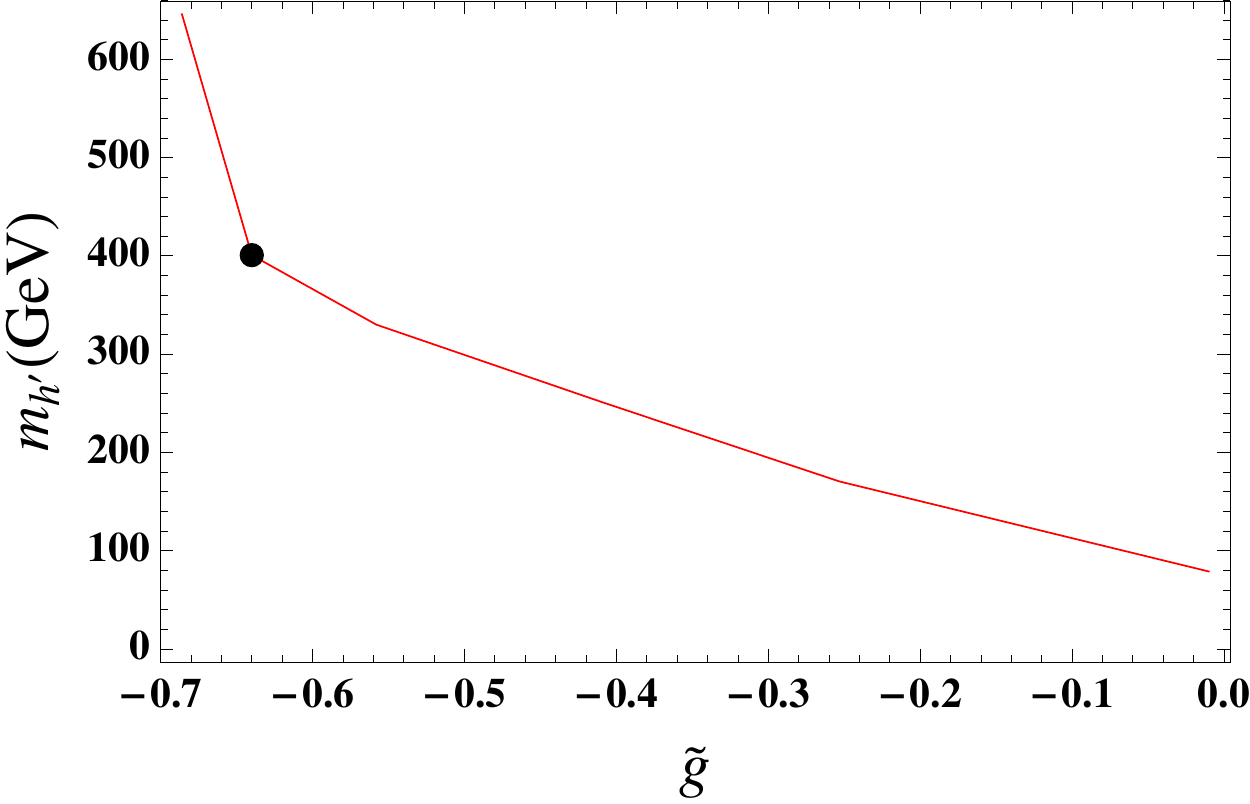}~\includegraphics[scale=.40]{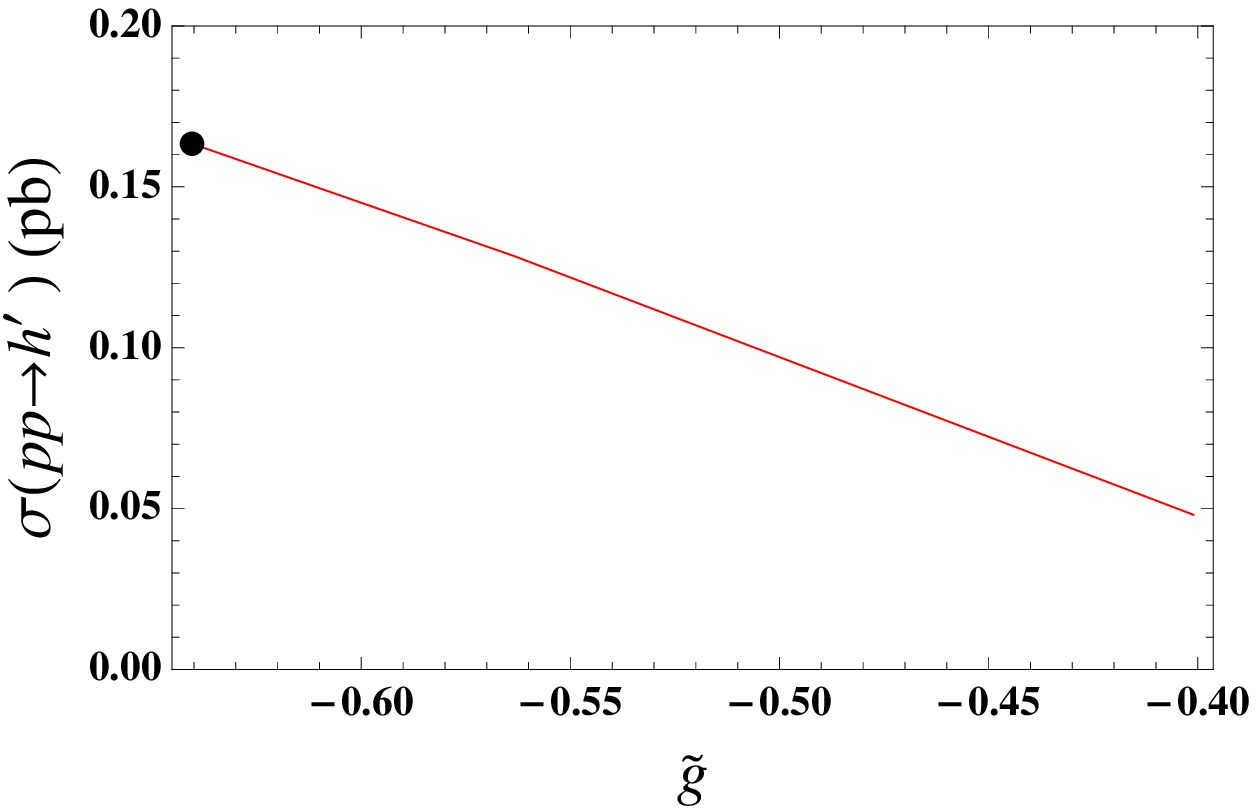}~\includegraphics[scale=.55]{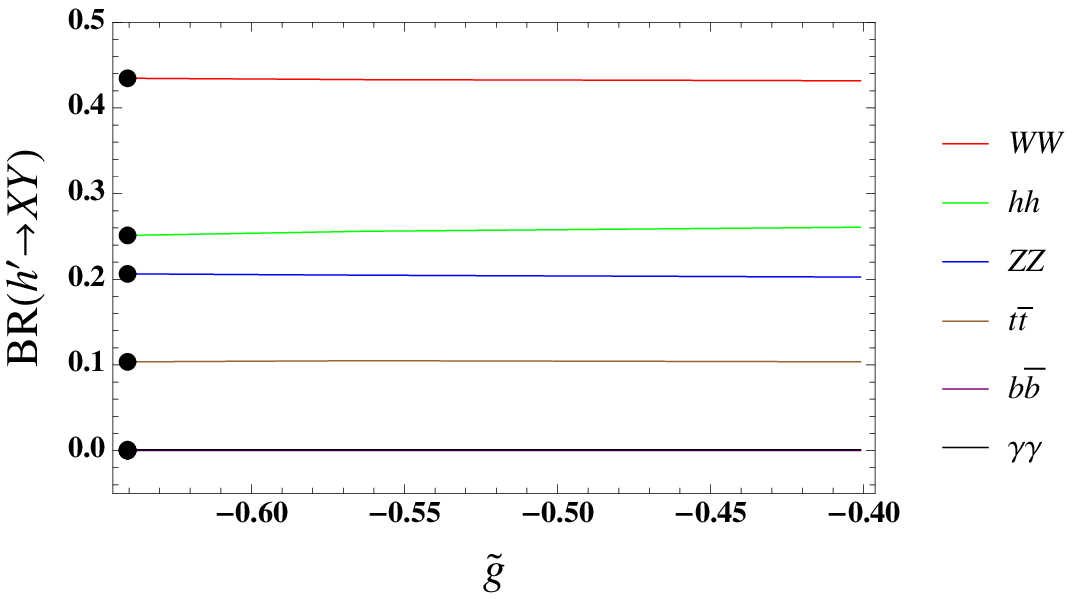}
\caption{\label{Fig:ggFBR}The dependence of (left) $m_{h'}$, (middle) the $h'$ production cross section via ggF at $\sqrt{s}=14~\text{TeV}$ and (right) $h'$ decay BRs (right) upon the gauge kinetic mixing coupling $\tilde{g}$. The values corresponding to the BP of Table~\ref{Tab:BPPar} are labeled by $\bullet$.}
\end{figure}

In Fig.~\ref{Fig:ggFBR} (left), we fix the SM-like Higgs boson mass to its measured value, i.e., $m_h\sim125~\text{GeV}$, and show the change of $m_{h'}$ with the gauge kinetic mixing parameter $\tilde{g}$. However, one should be careful when reading this panel, as we only chose to show $\tilde{g}$ and corresponding $m_{h'}$ values that give maximal values to the MSSM and $B-L$ Higgs sectors mixing represented in $Z^H_{22}$ shown in Fig.~\ref{Fig:mhpgtgbl}, as desired for our study, where all BPs are validated by~\Higgsbounds~and~\Higgssignals. Generally, this $m_{h'}-\tilde{g}$ subfigure would instead show a scattered pattern, as other BLSSM parameters could be tweaked such that any value of $\tilde{g}$ can correspond to a broad range of $m_{h'}$. The cross section for ggF, properly convoluted with the default Parton Distribution Functions (PDFs) of our ME generator (namely, $\sigma(pp\to h')$), as function of $\tilde g$, is found in Fig.~\ref{Fig:ggFBR} (middle), for $\sqrt s=14$ TeV. 
Also, in Fig.~\ref{Fig:ggFBR}~(right) we show the $h'$ decay BRs, again, as functions of $\tilde g$. In all three plots, the symbol
$\bullet$ refers to the BP adopted here, for which the corresponding $\sigma$ and BR values are found in Table~\ref{Tab:hpprdcyxsec}. The production cross section of $h'$ depends significantly on $\tilde g$, which is (as mentioned) the only source of mixing between the BLSSM Higgs $\chi_{1,2}$ singlets and the MSSM Higgs doublets $H_{u,d}$ that enables $h'$ couplings with SM particles. However, the $h'$ decay BRs are not significantly affected by it because both the partial and total decay widths of $h'$ in each channel receive nearly the same contribution from $\tilde{g}$, which cancels out from the BRs. It is noteworthy that the three most significant decay channels are the bosonic ones in $W^+W^-$, $ZZ$ and $hh$. In contrast, the fermionic decay channels into $t\bar t$ and $b\bar b$ are relatively less significant. Therefore, in the forthcoming MC analysis, we will concentrate on the former three decay channels.

For each channel, there are many corresponding background processes and all can be reduced by applying the cut-flows of Tables~\ref{Tab:sgnbkgnevntww2lmet},~\ref{Tab:sgnbkgnevntzz4l}~and~\ref{Tab:sgnbkgnevnthhbbaa}, in correspondence of the three aforementioned channels, respectively. What remain in all cases, though, are the irreducible backgrounds $pp\to{2\ell}+\slashed{E}_T$, $pp\to{4\ell}$ and $pp\to{\gamma\gamma{b\bar{b}}}$.
The following standard acceptance cuts on transverse momentum $(P_T)$, pseudorapidity $(\eta)$ and angular separation $(\Delta{R})$ of the final state leptons, jets and photons are applied: $(P_T)_{j}\geq20,~(P_T)_{a}\geq10,~|\eta_{j}|\leq5,~|\eta_{a}|\leq2.5,~a=\gamma,\ell$ and $\Delta{R}_{ab}\geq0.4,~a,b=j,\gamma,\ell$.

In Tables~\ref{Tab:sgnbkgnevntww2lmet},~\ref{Tab:sgnbkgnevntzz4l}~and~\ref{Tab:sgnbkgnevnthhbbaa}, the kinematical variables are defined such that $M_{\text{eff}}$ is the effective mass being obtained as the sum of the transverse momentum of all final state objects and the transverse energy, while $E_T$ is the scalar sum of the transverse energy of all
(visible) final state objects in the plane transverse to the beam~\cite{Conte:2012fm}. Furthermore, $M_{ab\ldots}$ is an invariant mass and $\Delta{R}_{ab}$ is the separation between final state objects. (Note that an (opposite-sign) di-lepton mass reconstruction around one $M_Z$ value in the $4\ell$ channel is not useful, as the irreducible background is here dominated by $pp\to ZZ,Z\gamma^*\to 4\ell$.)
\begin{table}[ht]
\centering
\begin{tabular}{lc}
\hline\hline
Quantity & Value \\
\hline
$\text{BR}(h'\to W^+W^-)$ 	& $0.432$ \\
$\text{BR}(h'\to ZZ)$ 	& $0.203$ \\
$\text{BR}(h'\to{hh})$ 	& $0.261$ \\
$\sigma(pp\to h')$ 		& $163.400~\text{(fb)}$ \\
$\sigma(pp\to h'\to W^+W^-\to 2\ell+\slashed{E}_T)$ 	&	$~~9.256~\text{(fb)}$ \\
$\sigma(pp\to h'\to ZZ\to4\ell)$ 						&	$~~0.406~\text{(fb)}$ \\
$\sigma(pp\to h'\to{hh}\to\bar{b}b\gamma\gamma)$		&	$~~0.124~\text{(fb)}$ \\
\hline\hline
\end{tabular}
\caption{\label{Tab:hpprdcyxsec}Production cross section $\sigma$ (at $\sqrt{s}=14~\text{TeV}$) and decay BRs into $W^+W^-$,
$ZZ$ and $hh$
for the $h'$ state 
(with $m_{h'}=400~\text{GeV}$) of our BP, including the overall rates in the three final states $2\ell+\slashed{E}_T$, $4\ell$ and $b\bar b\gamma\gamma$. Normalization is to LO for all $\sigma$'s.}
\end{table}

\begin{table}[h]
\begin{subtable}{1.\textwidth}
\captionsetup{justification=centering}\centering
{
\begin{tabular}{lllll}
\hline\hline
Cuts (select) 							&& $S$ 				& $B$								& $S/\sqrt{B}$			\\\hline
Initial (no cut) 						&& $6120~~~~~$ 		& $8913654030$						& $0.065$				\\
ID and Kin cuts 						&& $2640^{\pm39}$ 	& $~~257832804^{\pm14207}$			& $0.164^{\pm0.000}$	\\\
$E_T~~~~~>700~\text{GeV}$				&& $~~338^{\pm18}$	& $\quad\quad~~ 70557^{\pm256}$		& $1.272^{\pm0.004}$	\\\ 
$M_{T}^{\ell^+\ell^-}>115~\text{GeV}$ 	&& $~~267^{\pm16}$	& $\quad\quad~~~~ 3171^{\pm56}$		& $4.734^{\pm0.004}$	\\
\hline\hline
\end{tabular}
\caption{\label{Tab:sgnbkgnevntww2lmet}$pp\to h'\to W^+W^-\to 2\ell+\slashed{E}_T$ cut-flow at $L_{\text{int}}=300~\text{fb}^{-1}$.}
}
\end{subtable}
\\\vspace{0.3cm}
\begin{subtable}{1.0\linewidth}
\captionsetup{justification=centering}\centering
{
\begin{tabular}{lllll}
\hline\hline
Cuts (select) 						&& $S$				& $B$ 				& $S/\sqrt{B}$ \\\hline 
Initial (no cut)					&& $267$			& $9712$ 			& $2.72$ 		\\
$E_T~~~~~>300~\text{GeV}$ 			&& $209^{\pm7}$	& $1680^{\pm37}$	& $5.09^{\pm0.004}$ \\
${M_{\ell^+\ell^-}}^{>~50~\text{GeV}}_{<150~\text{GeV}}$ 	&& $173^{\pm8}$	& $1395^{\pm35}$	& $4.63^{\pm0.005}$ \\
\hline\hline
\end{tabular}
\caption{\label{Tab:sgnbkgnevntzz4l}$pp\to h'\to{ZZ}\to4\ell$ cut-flow at $L_{\text{int}}=300~\text{fb}^{-1}$.}
}
\end{subtable}
\\\vspace{0.3cm}
\begin{subtable}{1.\linewidth}
\captionsetup{justification=centering}\centering
{
\begin{tabular}{lrrll}
\hline\hline
Cuts (select) 								&& $S~\quad\quad$& $B$					& $S/\sqrt{B}$ \\\hline
Initial (no cut) 							&& $951~\quad$ 	& $19951560$			& $0.213$\\
$E_T~~>200~\text{GeV}$ 					&& $933^{\pm~4}$& $~~1476867^{\pm1169}$	& $0.768^{\pm0.000}$\\
${M_{\gamma\gamma}}^{~>120~\text{GeV}}_{~<135~\text{GeV}}$ 			&& $475^{\pm15}$& $~~~~~~29023^{\pm170}$& $2.787^{\pm0.001}$\\
${M_{bb}}^{~~>~50~\text{GeV}}_{~~<160~\text{GeV}}$ 					&& $135^{\pm11}$& $~~~~~~~~1945^{\pm44}$& $3.055^{\pm0.005}$\\
${\Delta R}^{\gamma\gamma<3.5}_{b\bar{b}~<3.5}$					&& $132^{\pm11}$& $~~~~~~~~1746^{\pm42}$& $3.156^{\pm0.006}$\\
${M_{\gamma\gamma\bar{b}b}}^{>360~\text{GeV}}_{<450~\text{GeV}}$	&& $~~99^{\pm10}$& $~~~~~~~~~~403^{\pm20}$& $4.903^{\pm0.017}$\\
\hline\hline
\end{tabular}
\caption{\label{Tab:sgnbkgnevnthhbbaa}$pp\to h'\to{hh}\to{b\bar{b}\gamma\gamma}$ cut-flow at $L_{\text{int}}=3000~\text{fb}^{-1}$.}
}
\end{subtable}
\caption{\label{tab:BP1}$S$ vs $B$ rates for the three signals pursued in our analysis in correspondence of our BP: the $2\ell+\slashed{E}_T$ (a), $4\ell$ and $b\bar b\gamma\gamma$ (c) final state. We adopt here $\sqrt{s}=14~\text{TeV}$
and integrated luminosity of Run 3 and HL-LHC. Inclusive NNLO $k$-factors from QCD are used here throughout.}
\end{table}

\subsection{THE $h'\to W^+W^-\to 2\ell+\slashed{E}_T$ CHANNEL}
\begin{figure}[ht]
\centering
\includegraphics[scale=0.55]{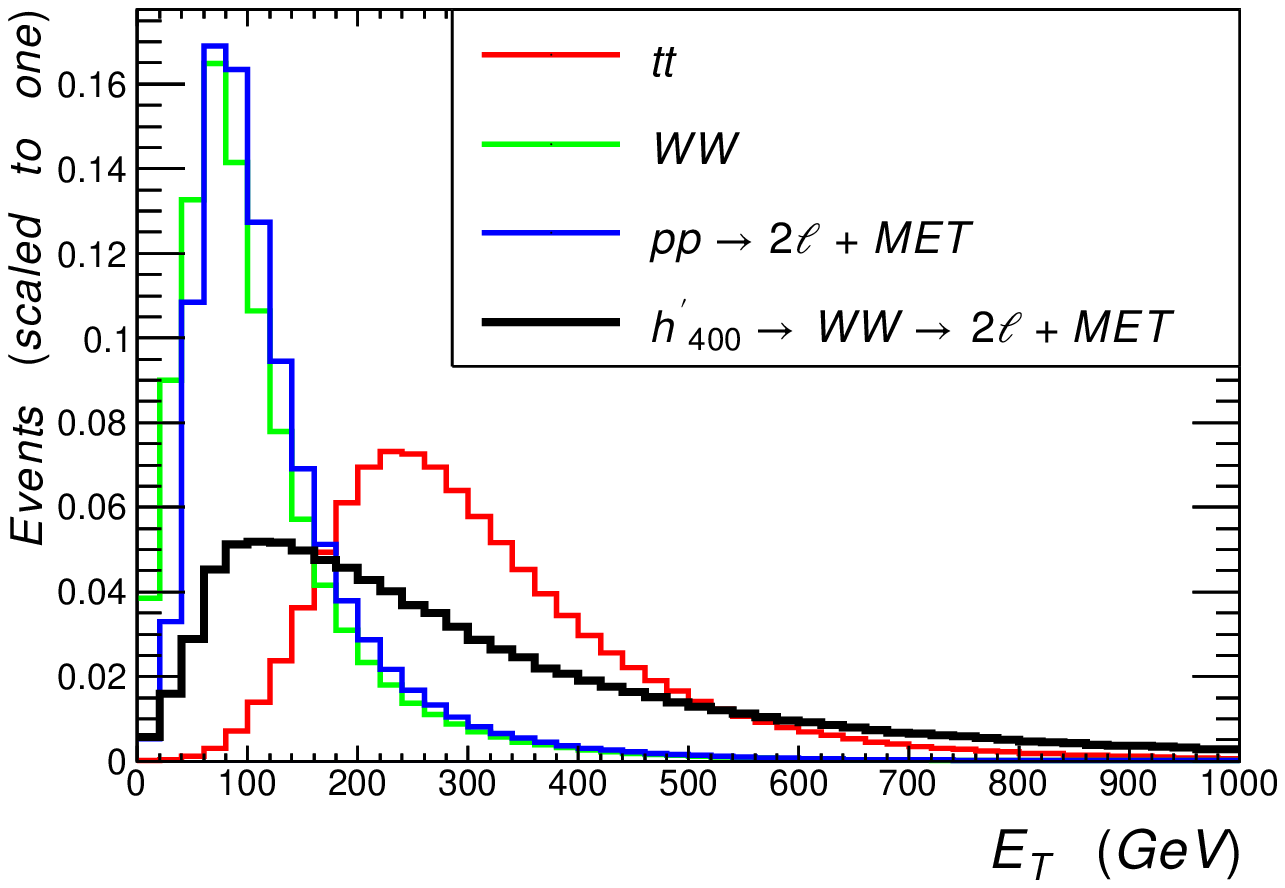}~\hspace{-1in}~\includegraphics[scale=0.55]{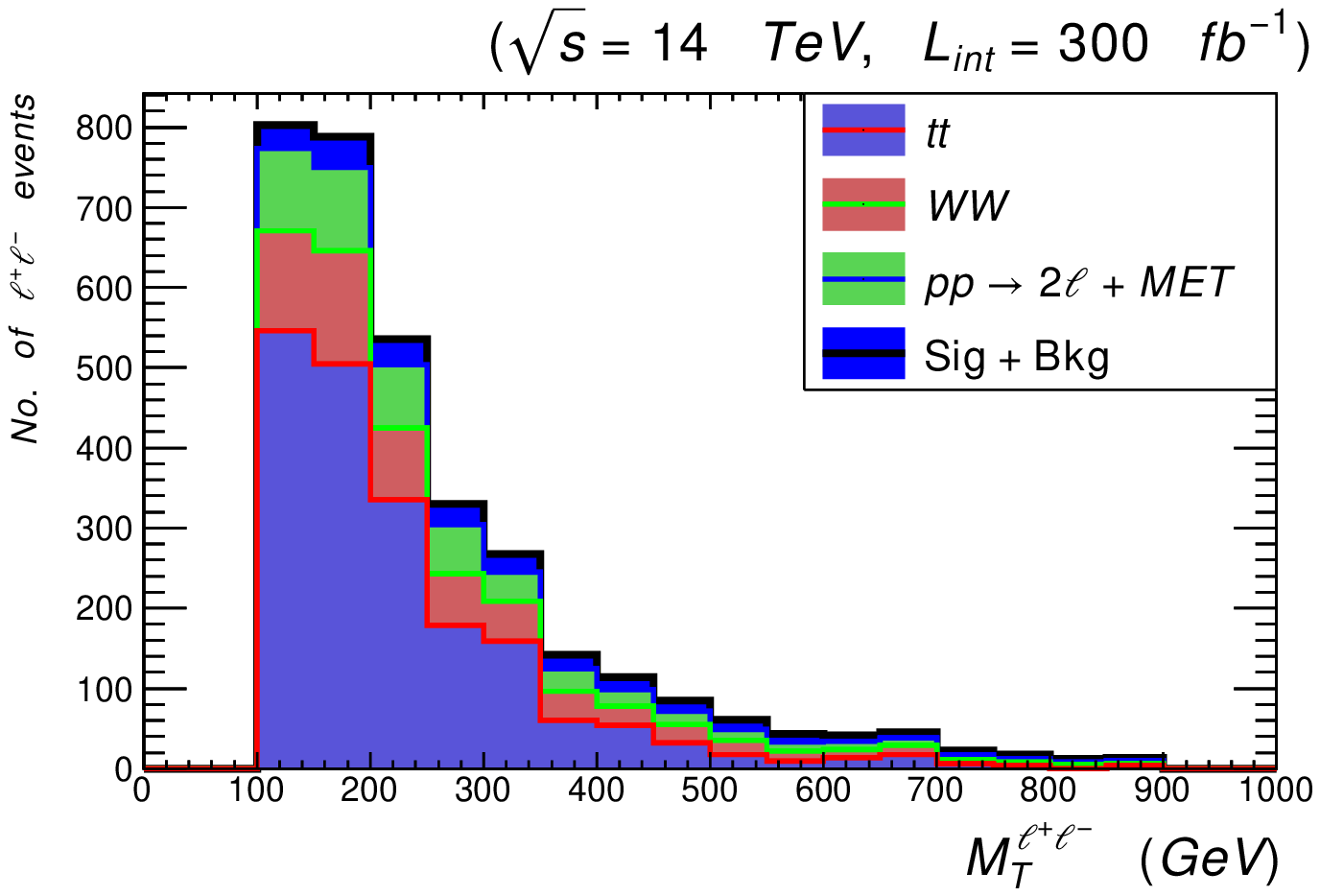}
\caption{\label{Fig:HWW2lMET} $S$ and $B$ distributions in $E_T$ normalized to 1 before applying the cut-flow (left), and stacked normalized to the total event rate $M_{T}^{\ell^+\ell^-}$ after applying the cut-flow (right) for the integrated luminosity $L_{\text{int}}=300~\text{fb}^{-1}$. In both cases we show both the ggF contribution to the signal $tt$ (red), $WW$ (green), $2\ell+\slashed{E}_T$ (blue) backgrounds and for our BP signal (black).} 
\end{figure}

Table~\ref{Tab:sgnbkgnevntww2lmet} provides the cut-flow for the $h'$ production and decay analysis via the $2\ell+\slashed{E}_T$ signature, while event shapes and rates (the latter in correspondence to Run 3 luminosity) for 
{
\begin{equation}\label{Eq:ppwwtotxs}
\sigma(pp\to h'\to W^+W^-\to 2\ell+\slashed{E}_T)\approx\sigma(pp\to h')\times\text{BR}(h'\to W^+W^-\to 2\ell+\slashed{E}_T)
\end{equation}
}
\noindent
are presented in Fig.~\ref{Fig:HWW2lMET}. Herein, we also present the contributions of an additional signal channel, induced by 
($W^+W^-$ dominated) VBF with two additional (untagged) forward/backward jets, as it contributes not negligibly to the same ggF signal regions (so that it has been taken into account in extracting our final sensitivities). In this figure,
the normalized (to 1) distributions used for the cut-flow (i.e., $E_T$, $M_{\text{eff}}$ and $\Delta R_{\ell^+\ell^-}$) are presented, alongside the full transverse mass {($M_{T}^{\ell^+\ell^-}=\sqrt{(E_T^{\ell\ell}+\slashed{E}_T)^2+|\vec{P}_T^{\ell\ell}+\vec{\slashed{E}}_T|^2}$, where $E_T^{\ell\ell}=\sqrt{|\vec{P}_T^{\ell\ell}|^2+m_{\ell\ell}^2}$, and $\vec{\slashed{E}}_T$ is the negative vector sum of the transverse momenta of the reconstructed objects, including
muons, electrons, photons, jets)} of the final state (i.e., using both leptons in its definition), the latter 
integrating to the actual event numbers for Run 3 and also in presence of the background contribution. Altogether, from this last spectrum,
it is clear that a high signal significance can be reached, however, it also shows that the shape does not promptly correlate to the $h'$ mass value. Yet, the significant excess seen in this channel will clearly motivate a parallel search in the $4\ell$ final state, which
we are illustrating in the next subsection. However, before doing so, let us dwell more on the noise composition. 

{
The dominant backgrounds in this channel are non-resonant $W^+W^-,~t\bar{t}$, and $W^\pm t$ production, all of
which have real $W^+W^-$ pairs in the final state. Other important backgrounds include Drell-Yan (DY) events ($pp\to Z/\gamma^{(*)}\to\ell^+\ell^-$) with $\slashed{E}_T$ that may arise from mis-measurements, $W^\pm+$ jets events in which a jet produces an object reconstructed as the second electron and $W^\pm \gamma$ events in which the photon undergoes a conversion. Boson pair production $W^\pm\gamma^*/~W^\pm Z^{(*)}/~Wh^{(*)}$ and $ZZ^{(*)}$ can also produce opposite-charge lepton pairs with additional leptons that are not detected.

Demanding the following set of identification cuts (ID) with the number of $b$-jets $N(b)<1$, the number of charged lepton pairs $N(\ell^+\ell^-)\leq2$ and the number of jets $N(j)\leq4$ in the kinematical (Kin) regions 
\begin{enumerate}
\item for the leading lepton $P^T_{\ell}\geq25$,
\item for the subleading lepton $P^T_{\ell}\geq15$ and 
\item for the two lepton $|\eta|_{\ell}<2.5$
\end{enumerate}
increases the $S$ to $B$ significance by a factor of about $2.5$. The final analysis is included in Table~\ref{Tab:sgnbkgnevntww2lmet}. After ID and Kin cuts, the DY, $W^\pm + $ jets, $W^\pm \gamma^{(*)}/Z^{(*)}$, $ZZ^{(*)}$ noises were eliminated so that in the end we kept only the irreducible backgrounds from $W^+W^-,~t\bar t$ and $pp\to2\ell+\slashed{E}_T$ events, which we stacked on top of each other in Fig.~\ref{Fig:HWW2lMET}.
}
%

\subsection{THE $h'\to{ZZ}\to4\ell$ CHANNEL}
\begin{figure}[ht]
\centering
\includegraphics[scale=0.55]{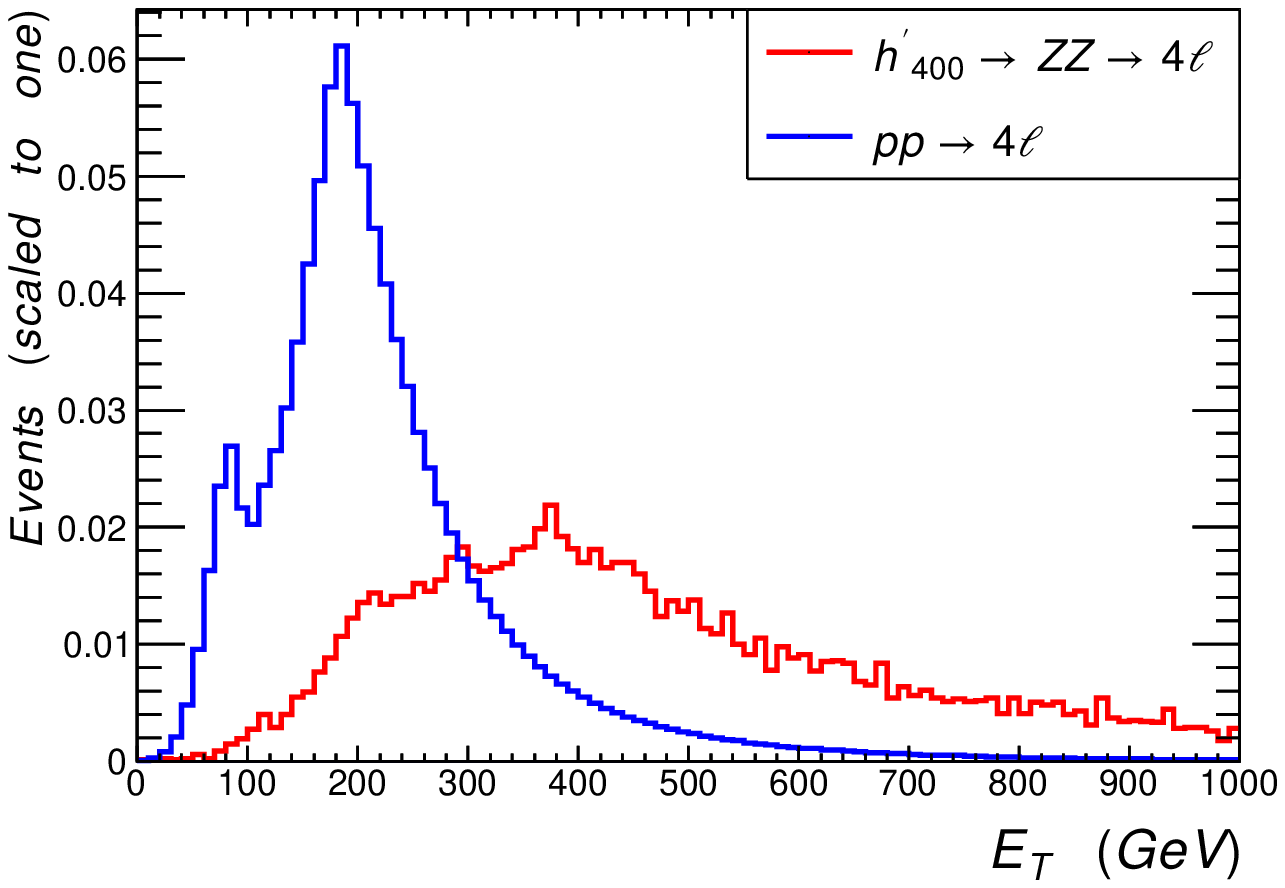}~\hspace{-1in}~\includegraphics[scale=0.55]{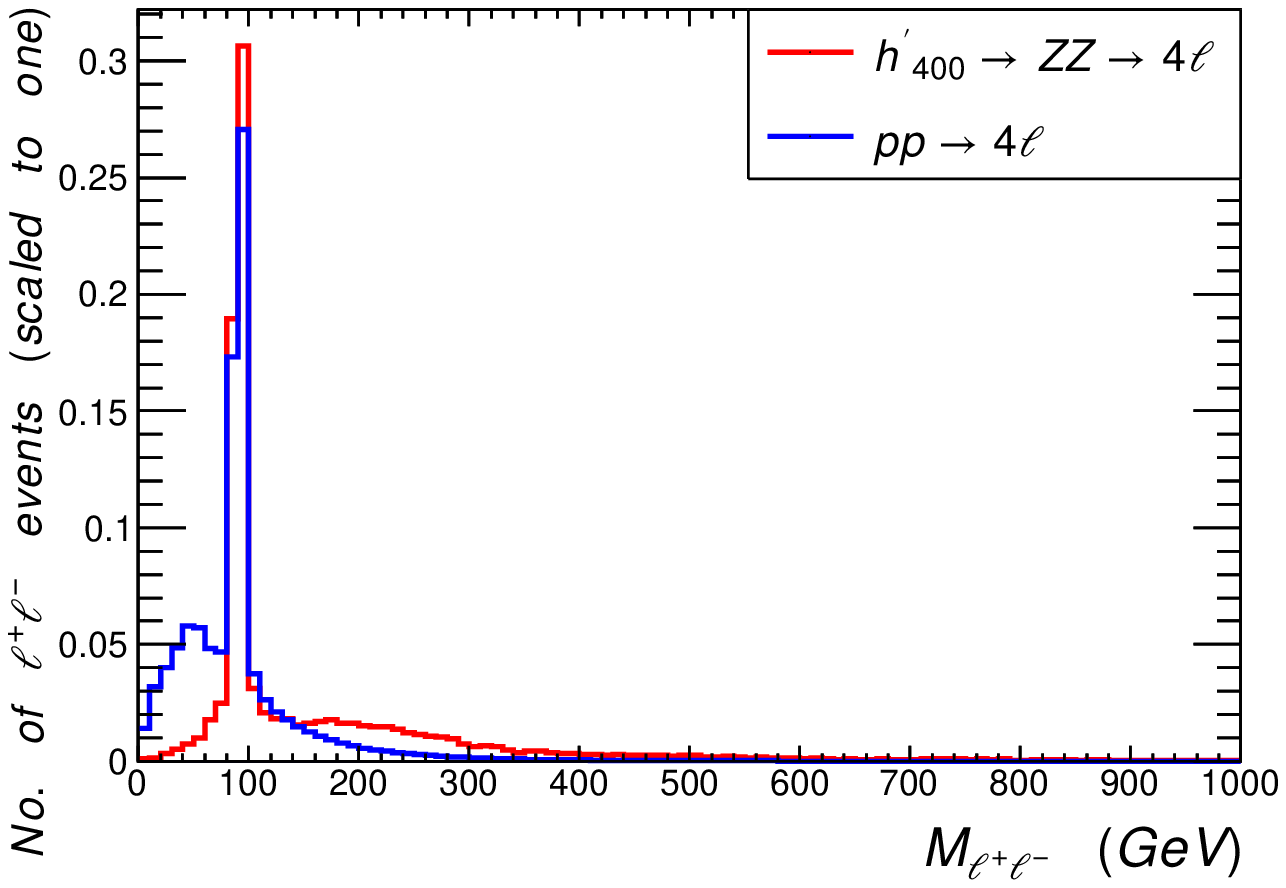}\\~\\
\hspace{1in}~\includegraphics[scale=0.55]{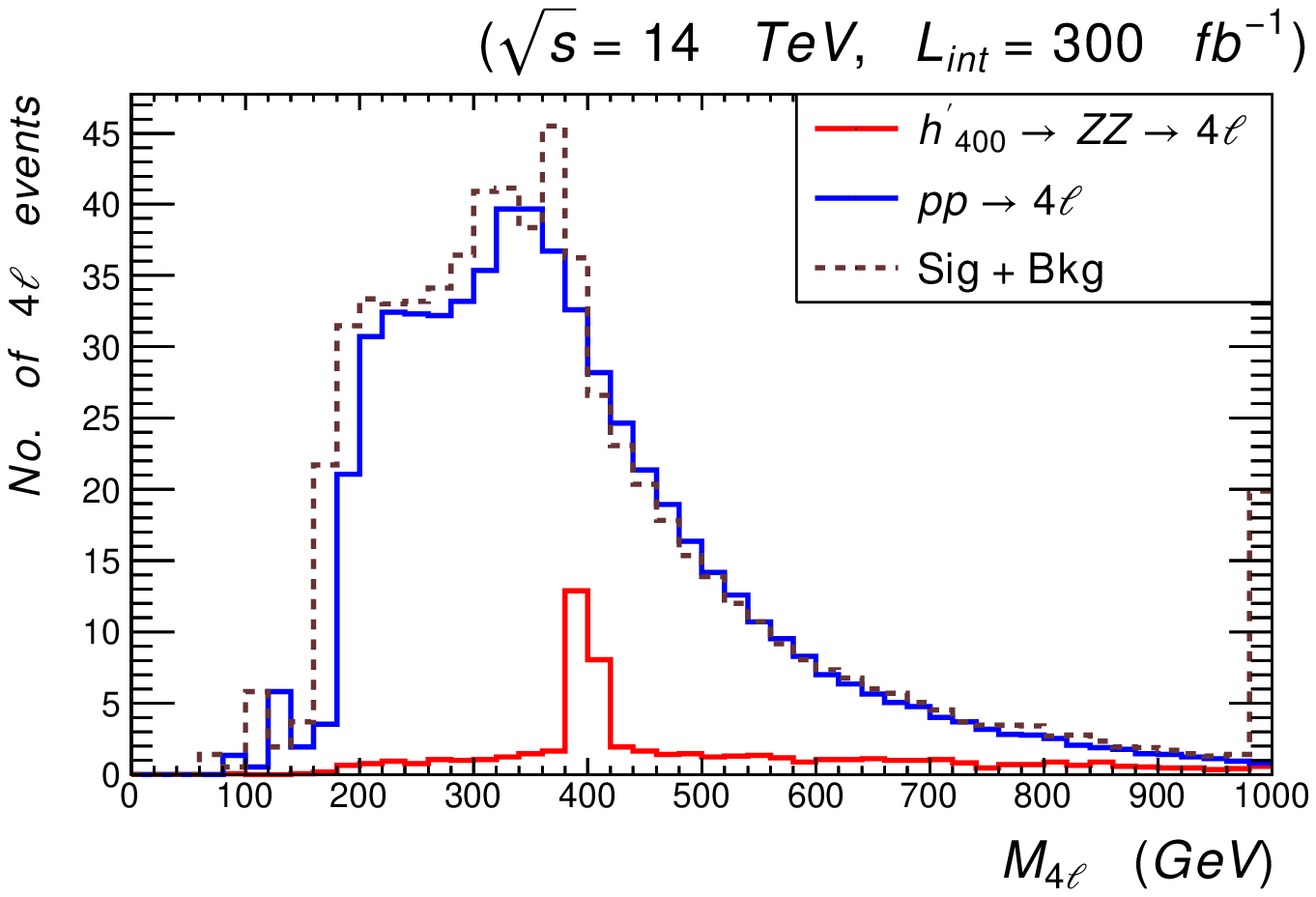}
\caption{\label{Fig:HZZ4l} $S$ and $B$ distributions in $E_T$ (top-left), $M_{\ell^+\ell^-}$ (top-right) and $M_{4\ell}$ (bottom), as defined in the text, the former two given before the cut-flow and normalized to 1 while the latter one given after it and normalized to the total event rate for the integrated luminosity $L_{\text{int}}=300~\text{fb}^{-1}$. In all cases we show only the ggF contribution to the 
$4\ell$ signal for our BP while for the last spectrum we also show the (stacked) distribution.}
\end{figure}

Table~\ref{Tab:sgnbkgnevntzz4l} provides the cut-flow for $h'$ production and decay via the $4\ell$ channel, while some 
relevant kinematics, in terms of event shapes and rates (the latter, again, in correspondence to Run 3 luminosity) for 
\begin{equation}\label{Eq:ppzztotxs}
\sigma(pp\to h'\to ZZ\to 4\ell)\approx\sigma(pp\to h')\times\text{BR}(h'\to ZZ\to 4\ell)
\end{equation}
\noindent
is presented in Fig.~\ref{Fig:HZZ4l}. Here, we concentrate on the normalized (to 1) distributions in transverse energy of all leptons ($E_T$) and opposite-sign di-lepton invariant mass ($M_{\ell^+\ell^-}$), both of which are used in our cut-flow. ({Regarding the latter, notice that the loss of significance in applying the cut in invariant mass against the dominant irreducible background $pp \to ZZ,Z\gamma^* \to4\ell$ is rather insignificant against the benefits of rejecting the irreducible one, e.g., from top-antitop quark production and fully leptonic $W^+W^-$ decays (which has typically a harder distribution in this variable), so that the whole of the latter can be neglected.}) In the end, the spectrum from which to extract the $h'$ resonance, i.e., the final state invariant mass, $M_{4\ell}$, clearly reveals a broad excess over a 400 GeV or so mass interval, altogether yielding significances in the discovery range. In fact, also a noticeable peak appear for $M_{4\ell}\approx400$ GeV (which, as mentioned, can be correlated with the $M_{T}^{\ell^+\ell^-}$ distribution in the $2\ell+\slashed{E}_T$ final state), so that one can improve further the potential for 
$h'$ discovery in the $4\ell$ channel by optimizing a cut in this variable. 

%

\subsection{THE $h'\to{hh}\to{b\bar{b}\gamma\gamma}$ CHANNEL}
\begin{figure}[ht]
\centering
\includegraphics[scale=0.55]{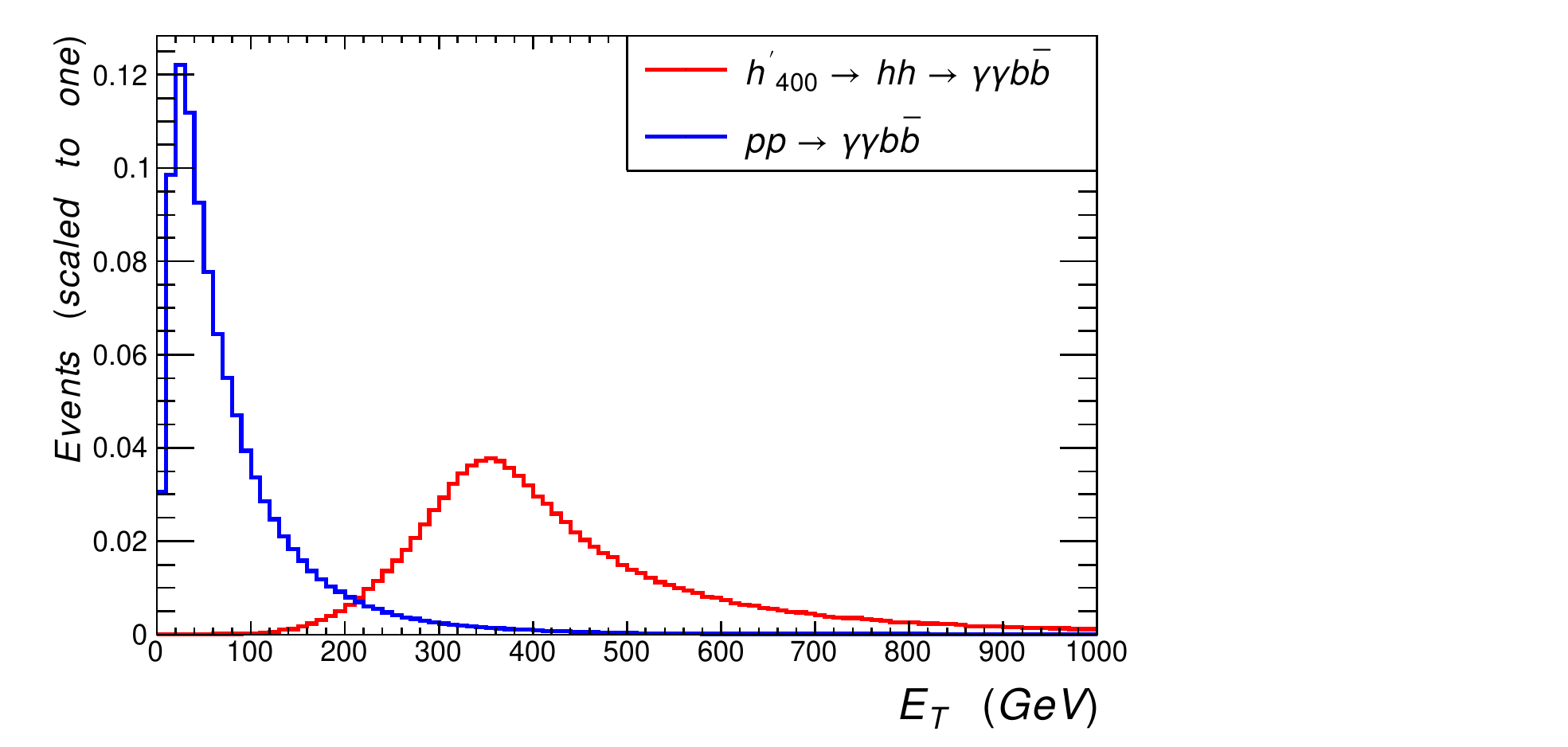}~\hspace{-1in}~\includegraphics[scale=0.55]{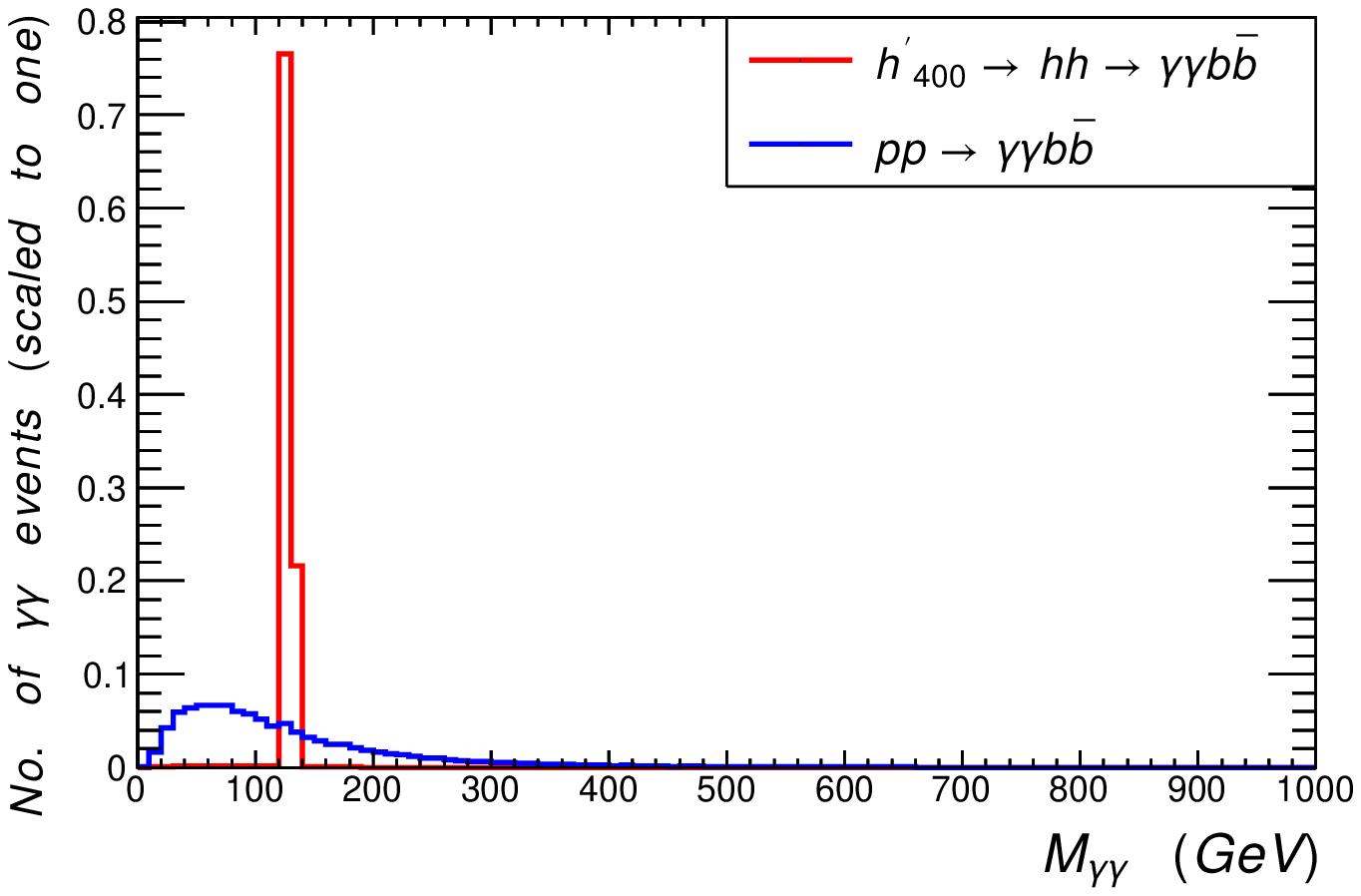}
\includegraphics[scale=0.55]{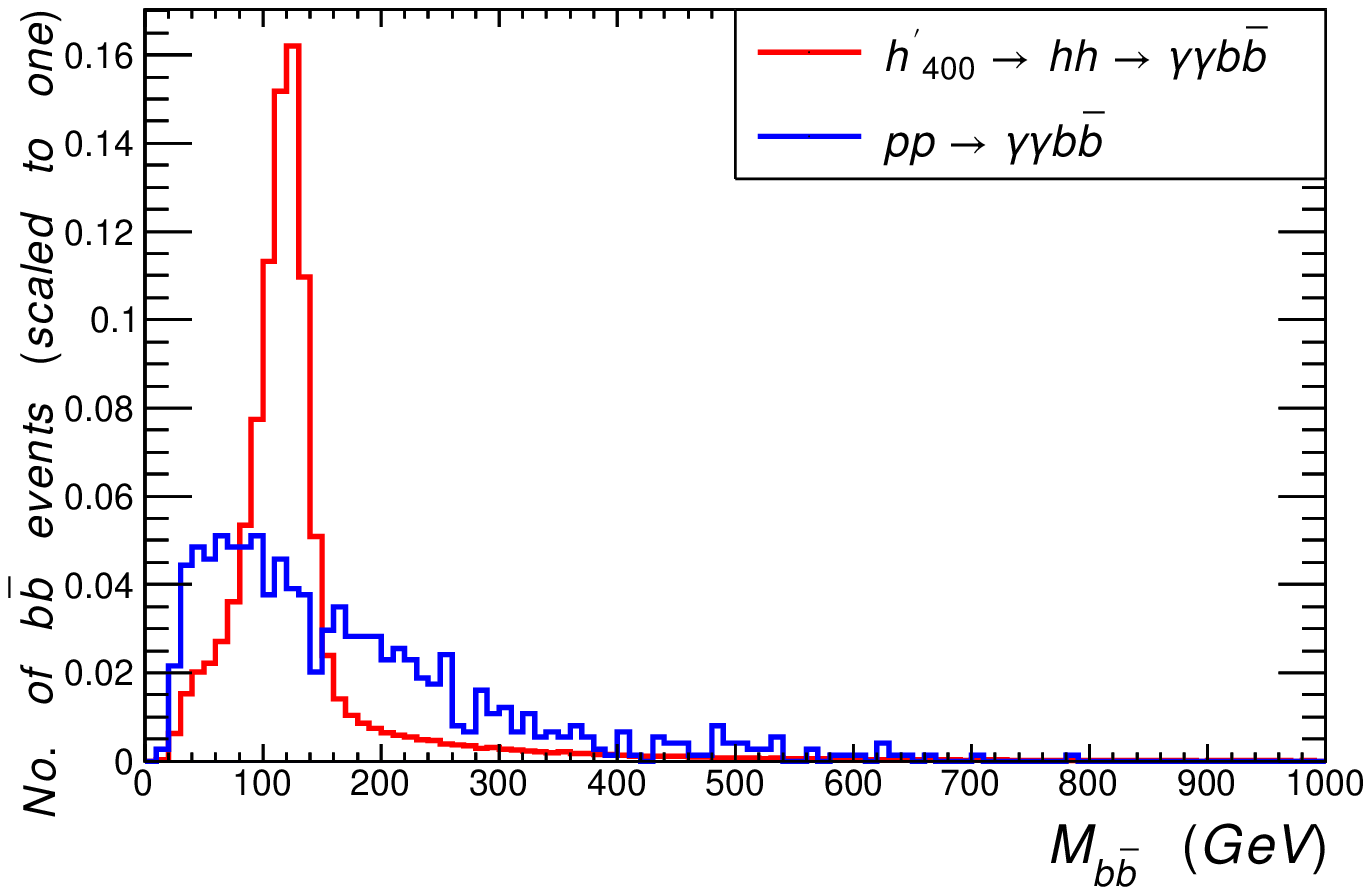}~\hspace{-1in}~\includegraphics[scale=0.55]{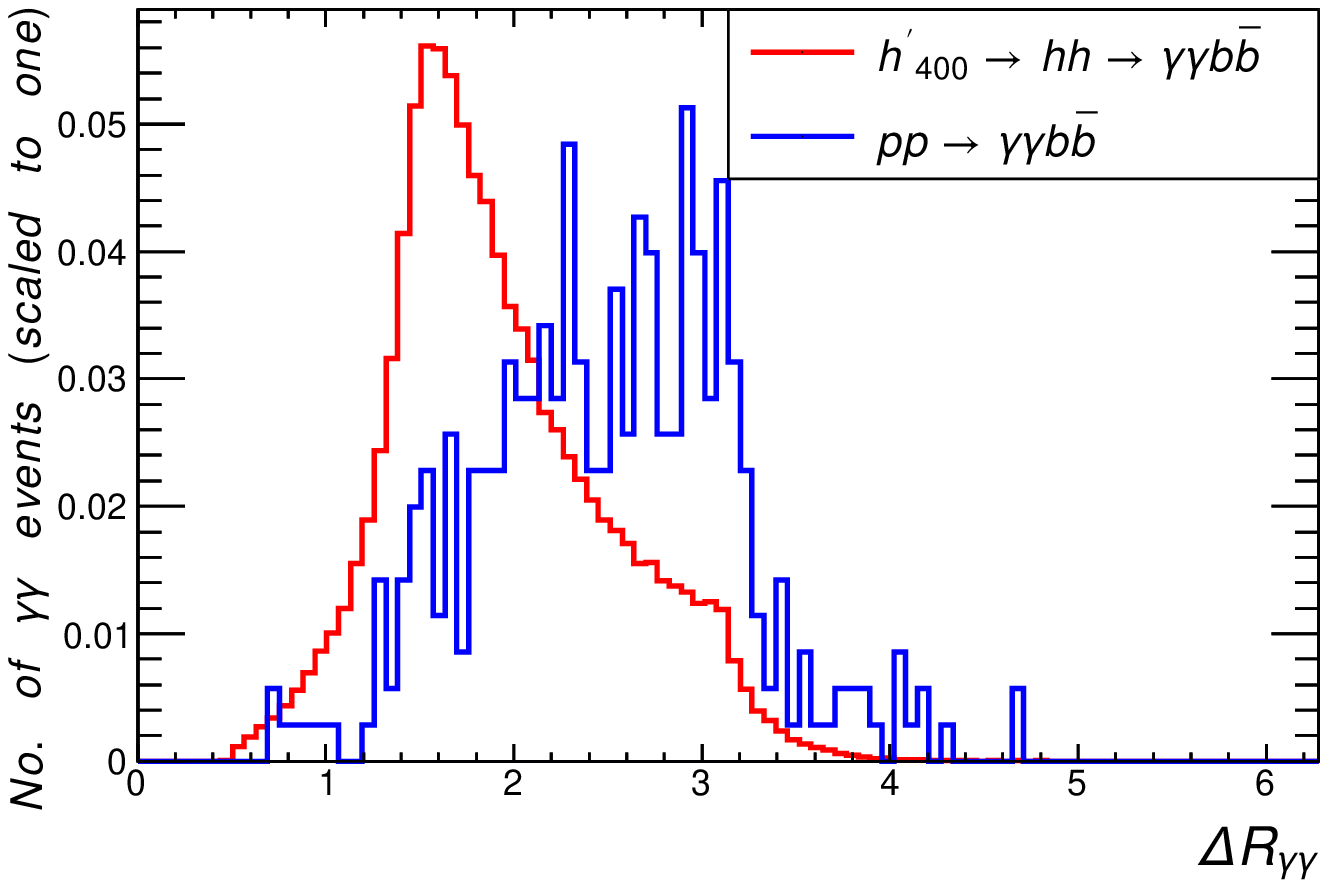}
\includegraphics[scale=0.55]{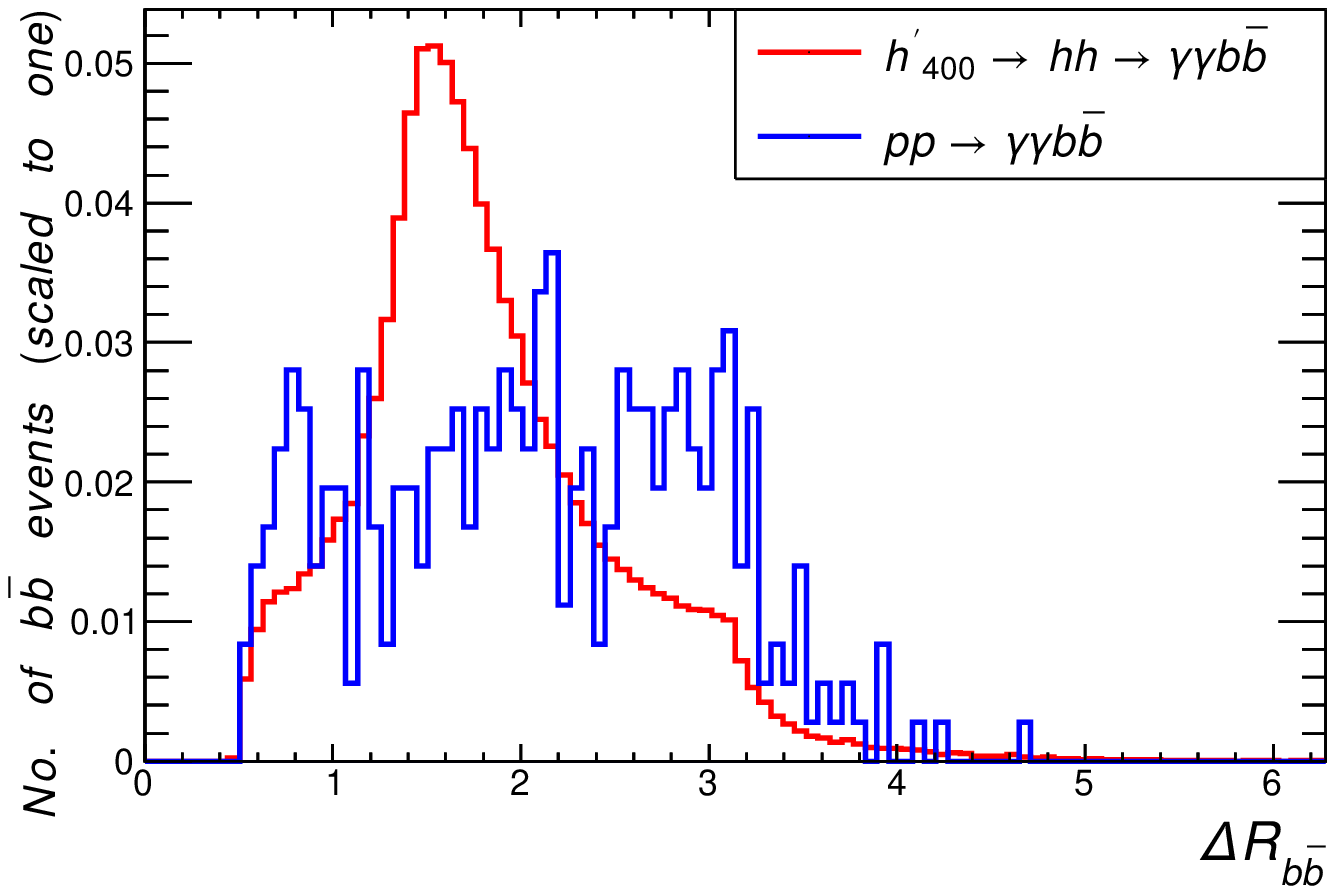}~\hspace{-1in}~\includegraphics[scale=0.55]{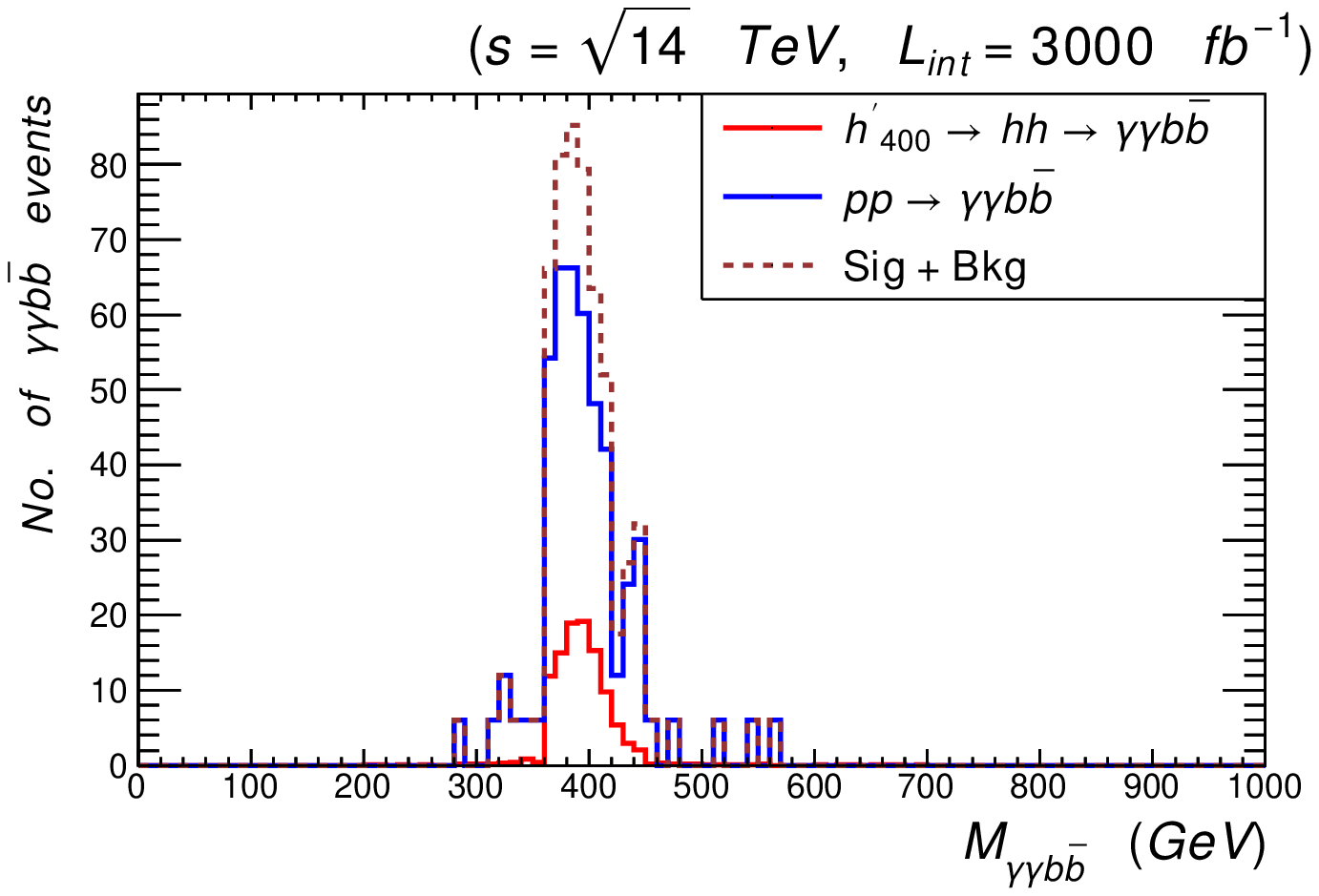}
\caption{\label{Fig:Hhhaabb} $S$ and $B$ distributions in $E_T$ (top-left), $M_{\gamma\gamma}$ (top-right), $M_{b\bar b}$ (middle-left),
$\Delta R_{\gamma\gamma}$ (middle-right)
$\Delta R_{b\bar b}$ (bottom-left) and 
$M_{\gamma\gamma b\bar b}$ (bottom-right), 
 as defined in the text, the former 5 given before the cut-flow and normalized to 1 while the latter one given after it and normalized to the total event rate for the integrated luminosity $L_{\text{int}}=3000~\text{fb}^{-1}$. In all cases we show only the ggF contribution to the 
$b\bar b\gamma\gamma$ signal for our BP while for the last spectrum we also show the (stacked) distribution.}
\end{figure}

Table~\ref{Tab:sgnbkgnevnthhbbaa} provides the cut-flow for the $h'$ production and decay analysis of the last channel we study, 
\begin{equation}\label{Eq:pphhtotxs}
\sigma(pp\to h'\to{hh}\to{b\bar{b}\gamma\gamma})\approx\sigma(pp\to h')\times\text{BR}(h'\to{hh}\to{b\bar{b}\gamma\gamma}),
\end{equation}
wherein we use HL-LHC luminosity, as this channel is not accessible during Run 3. The distributions used to inform our cut-flow herein (normalized to 1) are found in Fig.~\ref{Fig:Hhhaabb}. These are the spectra in 
the
transverse energy of the $b\bar b\gamma\gamma$ final state 
 ($E_T$), $\gamma\gamma$ and $b\bar b$ invariant masses ($M_{\gamma\gamma}$ and $M_{b\bar b}$, respectively) and separations ($\Delta R_{\gamma\gamma}$ and $\Delta R_{b\bar b}$, respectively). Such a figure also presents the invariant mass
of the final state ($M_{\gamma\gamma b\bar b}$), normalized to the HL-LHC luminosity. As seen from the
signal and background responses to the cut-flow, it is clear that knowledge of the $m_{h'}$ value, gained during Run 3 of the LHC by exploiting the two previous signatures, is crucial in accessing this signal, which can ultimately be done at the 5$\sigma$ level, despite the initially overwhelming background. 
\subsection{Historical Significances}
\begin{figure}[ht]
\centering
\includegraphics[scale=.5]{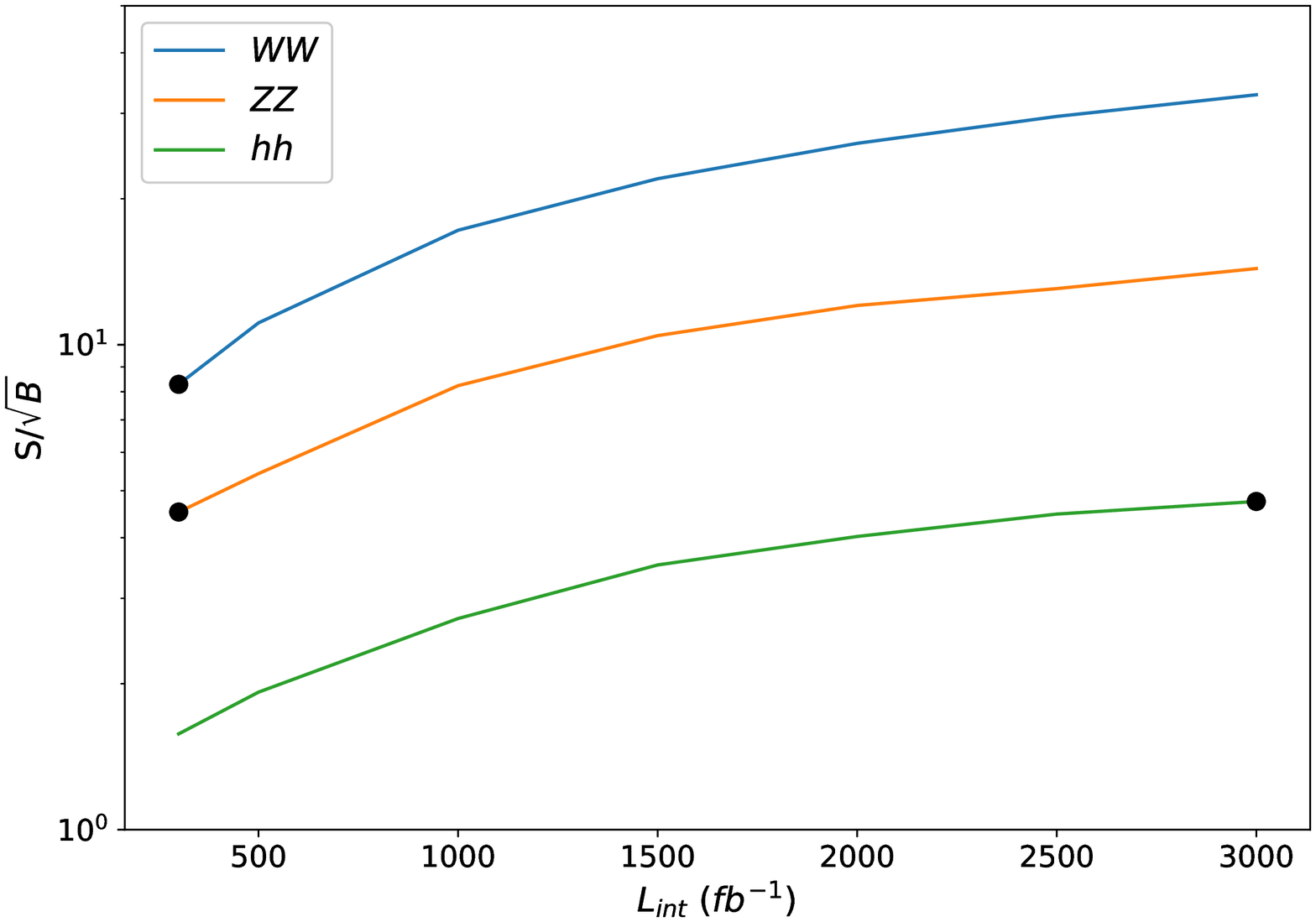}
\caption{\label{Fig:HVVSigLumi}Significance of the $h'\to W^+W^-,ZZ$ and $hh$ signals that we have studied versus $L_\text{int}$ for our BP. Data are produced at a center-of-mass energy of $\sqrt{s}=14~\text{TeV}$. The rates are computed after applying the relevant kinematical analyses described in the text. The three $\bullet$ points indicate the luminosity choices used in the MC simulations performed.} 
\end{figure}

Before closing this section, we describe the patterns of significances in the three channels that we have studied, 
as they would evolve with luminosity, assuming fixed energy at $\sqrt s=14$ TeV. These are shown in Fig.~\ref{Fig:HVVSigLumi}. 
lt is evident that a full characterization of the $h'$ state, involving its coupling to SM (massive) gauge and Higgs bosons is only possible through a combined effort of analyses to be entertained at both Run 3 of the LHC and HL-LHC. 

\section{\label{Sec:cnclsn}CONCLUSIONS}
In summary, we have shown that a theoretically well-motivated realization of supersymmetry, the so-called BLSSM, may yield detectable signals of a heavy
neutral CP-even Higgs boson at the LHC, both during Run 3 and the HL-LHC phase. These emerge from the lightest (neutral) Higgs state of this scenario with prevalent $B-L$ composition, $h'$, while the lightest (neutral) Higgs state with predominant MSSM nature is identified with the discovered one, $h$ (with $m_h=125~\text{GeV}$). The subprocesses pursued to this effect, assuming a BP with an illustrative mass $m_{h'}=400~\text{GeV}$, have been $gg\to h'\to W^+W^-\to 2\ell+\slashed{E}_T$, $gg\to h'\to{ZZ}\to4\ell$ and $gg\to h'\to{hh}\to{b\bar{b}\gamma\gamma}$. The first one would be accessible during the early stages of Run 3 and the study of mass distributions would allow one to extract an indication of the $h'$ mass. This information can then be used to optimize the selection 
of the second signal, which would reveal a clear pick centered around $m_{h'}$ by the end of Run 3. With the latter information available, one would then be able to establish the third signal at the HL-LHC. All this will therefore enable one to fully characterize the $h'$ state, not only through its mass, but also in terms of its couplings, as the $W^+W^-$, $ZZ$ and $hh$ decays are the dominant ones in the BLSSM while those to $t\bar t$ and $b\bar b$ pairs may be accessible at production level through the ggF channel. This
finally opens up the possibility of eventually separating the BLSSM hypothesis from alternative ones also based on supersymmetry, since -- thanks to the peculiar feature of (gauge) kinetic mixing appearing in the BLSSM (which incorporates an additional $U(1)_{B-L}$ group beyond the
SM gauge symmetries) -- competing signals stemming from, e.g., the MSSM would have rather different mass and coupling patterns. 

We have come to these conclusions by performing a full MC analysis in presence of ME, PS, fragmentation/hadronization effects as well as detector modeling and upon devising dedicated cut-and-count cut-flows for each signature pursued. We are therefore confident that ATLAS and CMS would have sensitivity to this specific non-minimal realization of supersymmetry and advocate dedicate searches for the aforementioned signals.

\section*{Acknowledgments} 
The work of M.A. is partially supported by the Science, Technology \& Innovation Funding Authority (STDF) under Grant No. 33495. The work of S.K. is partially supported by STDF under Grant No. 48173.
S.M. is supported in part through the NExT Institute and the Science \& Technology Facilities Council (STFC) Consolidated Grant No. ST/L000296/1.

\bibliographystyle{apsrev}
\bibliography{bib}
\end{document}